\documentclass[aps,letterpaper,prl,twocolumn,notitlepage,showpacs,superscriptaddress,
    nobalancelastpage,amsmath]{revtex4-1}
\usepackage{bbm,color,amsmath,amssymb,graphicx}
\usepackage[pdftex,
            pdfauthor={Yang-Le Wu},
            pdftitle={Quasihole}]{hyperref}

\begin{document}

\title{Braiding non-Abelian quasiholes in fractional quantum Hall states}

\author{Yang-Le Wu}
\affiliation{Department of Physics, Princeton University, Princeton, New Jersey 08544, USA}
\author{B. Estienne}
\affiliation{Sorbonne Universit\'es, UPMC Univ Paris 06, UMR 7589, LPTHE, F-75005, Paris, France}
\affiliation{CNRS, UMR 7589, LPTHE, F-75005, Paris, France}
\author{N. Regnault}
\affiliation{Department of Physics, Princeton University, Princeton, New Jersey 08544, USA}
\affiliation{Laboratoire Pierre Aigrain, ENS-CNRS UMR 8551, Universit\'es P. et M. Curie and Paris-Diderot, 24, rue Lhomond, 75231 Paris Cedex 05, France}
\author{B. Andrei Bernevig}
\affiliation{Department of Physics, Princeton University, Princeton, New Jersey 08544, USA}

\date{\today}

\begin{abstract}
Quasiholes in certain fractional quantum Hall states are promising candidates for 
the experimental realization of non-Abelian anyons. They are assumed to be 
localized excitations, and to display non-Abelian statistics when sufficiently 
separated, but these properties have not been explicitly demonstrated
except for the Moore-Read state. In this work, we apply the newly developed 
matrix product state technique to examine these exotic excitations. For the 
Moore-Read and the $\mathbb{Z}_3$ Read-Rezayi states, we estimate the 
quasihole radii, and determine the correlation lengths associated with the 
exponential convergence of the braiding statistics. We provide the first 
microscopic verification for the Fibonacci nature of the $\mathbb{Z}_3$
Read-Rezayi quasiholes.
We also present evidence for the failure of plasma screening in the 
nonunitary Gaffnian wave function.
\end{abstract}
\maketitle

Non-Abelian anyons have been the focus of much theoretical
and experimental interest due to the exciting prospect of topologically 
fault-tolerant quantum computing~\cite{Kitaev03:TQC,
Freedman02:Universal1,Freedman02:Universal2,Hormozi07:Compiling,Nayak08:RMP,
Levin05:StringNet,Fu08:Proximity,Lutchyn10:Majorana,Oreg10:Majorana,
Alicea11:Network,Alicea12:Majorana,Mourik12:Majorana,
Barkeshli11:Genon,Barkeshli12:Nematic,Barkeshli13:Twist,
Clarke13:FTSC,Lindner12:FTSC,Cheng12:FTSC,Vaezi13:FTSC,
Barkeshli13:Defects,Mong14:RR}.
As noted by Kitaev~\cite{Kitaev03:TQC},
the topological degeneracy of these exotic excitations allows nonlocal
storage of quantum information,
while adiabatic braiding implements unitary quantum gates.
Candidates~\cite{DasSarma05:TQC-MR,Stern06:TQC-MR,Bonderson06:TQC-RR}
for their physical realization are the quasiholes in
certain fractional quantum Hall states~\cite{Tsui82:FQH},
in particular, those around the plateaus at fillings
$\nu=5/2$ and $12/5$~\cite{Willett87:MR,Pan99:MR,Xia04:RR}.
Their model wave functions,
namely, the Moore-Read~\cite{Moore91:MR} (MR)
and the $\mathbb{Z}_3$ Read-Rezayi~\cite{Read99:RR} (RR) states,
enjoy an elegant first-quantized rewriting~\cite{Fubini91:FQH-CFT,Moore91:MR}
in terms of conformal field 
theory~\cite{Belavin84:BPZ,Moore89:CFT,DiFrancesco99:Yellow} (CFT) correlators,
from which many physical properties can be predicted.
The strengths of this approach rest on a crucial
\emph{conjecture}~\cite{Moore91:MR}:
quasihole braiding statistics can be directly read off from the monodromy 
of the CFT correlators.
Under this conjecture, the MR quasiholes are
Ising anyons, while the $\mathbb{Z}_3$ RR ones are Fibonacci anyons.
But the proof of the conjecture itself is lacking.

The relation between statistics and monodromy was originally established for the
Laughlin state~\cite{Laughlin83:Nobel}
through the plasma analogy~\cite{Arovas84:Statistics}.
Assuming sufficient quasihole separations,
the statistics-monodromy equivalence holds true when the plasma is in the
screening phase.
With considerable effort, this line of argument was recently extended to
the MR state~\cite{Gurarie97:Plasma,Read09:Adiabatic,Bonderson11:Plasma},
in agreement with finite-size
numerics~\cite{Tserkovnyak03:MR,Prodan09:Braiding,Baraban09:MR}.
More complicated states like the $\mathbb{Z}_3$ RR
still remain uncharted territory for both analytics and numerics,
despite their capacity for universal quantum
computation~\cite{Freedman02:Universal1,Freedman02:Universal2,Hormozi07:Compiling}.
Moreover, wave functions constructed from nonunitary field theories 
(such as the Gaffnian~\cite{Simon07:Gaffnian})
are conjectured not to give rise to sensible statistics~\cite{Read09:Adiabatic},
yet the microscopic symptom of such pathology is still under 
investigation~\cite{Estienne14:Gaffnian}.

\begin{table}[b]
\caption{\label{tab:quasihole-data}
Numerical data for quasihole radii $R$ and the correlation length 
$\xi_\mathrm{ortho}$
associated with wave function orthogonality
[see the discussion after Eq.~\eqref{eq:screening}].
}
\begin{ruledtabular}
\begin{tabular}{ccrrc}
& $\nu$ & \multicolumn{2}{c}{$R/\ell_0$} & $\xi_\text{ortho}/\ell_0$ \\[1pt]\hline
Laughlin & $\frac{1}{3}$ & \multicolumn{2}{c}{$\frac{e}{3}$: $2.6$} & - \\[2pt]\hline
Moore-Read & $\frac{1}{2}$ & $\frac{e}{4}$: $2.8$ & $\frac{e}{2}$: $2.7$ & $2.6$ \\[2pt]\hline
$\mathbb{Z}_3$ Read-Rezayi & $\frac{3}{5}$ & $\frac{e}{5}$: $3.0$ & $\frac{3e}{5}$: $2.8$ & $3.4$
\end{tabular}
\end{ruledtabular}
\end{table}

In this Letter, we aim to settle the aforementioned issues through numerical 
studies of certain model wave functions.
Until very recently, this was a daunting task due to the exponentially large 
Hilbert space, and in many cases, the absence of a convenient 
analytical form of the quasihole wave functions.
In fact, so far only the Laughlin and the MR quasiholes have been
tested directly, with various degrees of success,
using exact diagonalization and Monte Carlo 
techniques~\cite{Tserkovnyak03:MR,Prodan09:Braiding,Baraban09:MR}.
A similar check on the Gaffnian and the $\mathbb{Z}_3$ RR states proves
extremely challenging
due to the combinatorial complexity of their analytical expressions.
These difficulties are partially solved by the recent
development~\cite{Dubail12:MPS,Zaletel12:MPS,Estienne13:MPS,Estienne13:MPSLong}
of exact matrix product states~\cite{Fannes92:MPS,Schollwock11:DMRG-MPS}
(MPS)
for the CFT-derived wave functions~\cite{Fubini91:FQH-CFT,Moore91:MR}. 
The MPS formalism provides a faithful and efficient
representation of quantum Hall model states,
and greatly facilitates the calculation of physical observables.
We generalize this novel technique to non-Abelian quasiholes.
In the following, we focus on the physical results,
and leave the technical details to a forthcoming paper~\cite{Wu14:MPS}.
For the $\mathbb{Z}_{k\leq 3}$ RR states
(including Laughlin and MR at $k=1,2$), as well as the Gaffnian 
wave function,
we construct MPS for localized quasiholes, and estimate their radii from 
the electron density profile.
From adiabatic transport, we obtain the braiding matrices,
verify the link between statistics and monodromy for $\mathbb{Z}_{k\leq 3}$ 
RR at large quasihole separations,
and determine the associated length scales.
Numerical data are summarized in Table~\ref{tab:quasihole-data}.
Through these characterizations, we confirm the Fibonacci nature of the 
$\mathbb{Z}_3$ RR quasiholes, and we provide the first size 
estimate for these exotic excitations.
Our results also give a microscopic diagnosis for the pathology of
the Gaffnian.

We consider model wave functions defined by
CFT correlators~\cite{Fubini91:FQH-CFT,Moore91:MR}
on the cylinder geometry~\cite{Rezayi94:Cylinder}.
Evaluated in the Hamiltonian picture~\cite{Dubail12:MPS},
such conformal correlators allow a MPS representation over the Landau 
orbitals~\cite{Zaletel12:MPS},
with the auxiliary space being the truncated conformal Hilbert 
space~\cite{Yurov90:TCSA}.
This truncation is constrained by the entanglement area 
law~\cite{Eisert10:AreaLaw}: the auxiliary space has to grow exponentially 
with the cylinder perimeter $L_y$.
Compared with previous
studies~\cite{Dubail12:MPS,Zaletel12:MPS,Estienne13:MPS,Estienne13:MPSLong},
the new challenge here stems from the nontrivial fusion of non-Abelian 
quasiholes.
Fixing all the pinned quasihole coordinates does not specify a single wave 
function. Rather, it defines a multidimensional vector space of 
degenerate states~\cite{Moore91:MR,Nayak96:SO2n}.
A natural basis in this space arises from the conformal blocks~\cite{Moore91:MR},
labeled by topological charges on fusion tree 
diagrams~\cite{Moore89:CFT,Kitaev06:Anyon}.
Our construction produces a MPS for each conformal block.

\begin{figure}[]
\centering
\includegraphics[]{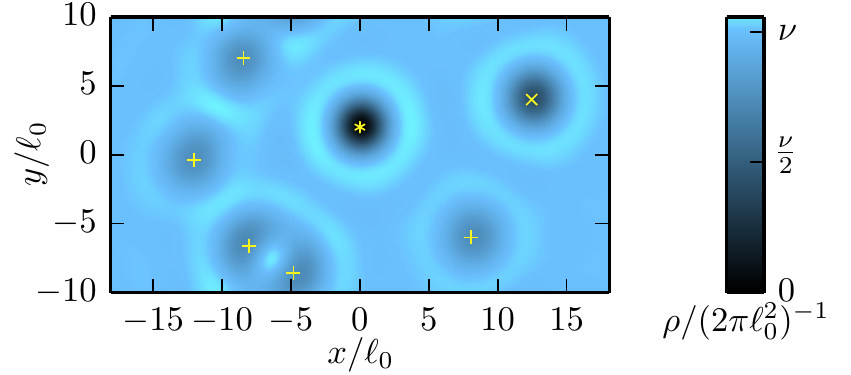}%
\caption{\label{fig:density-heatmap}
Electron density for $\mathbb{Z}_3$ RR
$\frac{e}{5}$($+$), $\frac{2e}{5}$($\times$) 
and $\frac{3e}{5}$($*$) quasiholes on an infinite cylinder with perimeter 
$L_y=20\ell_0$.
}
\end{figure}

We illustrate this procedure using the MR state.
It is constructed~\cite{Moore91:MR} from the chiral Ising
CFT~\footnote{We leave the compactified U(1) boson implicit.},
with primary fields
$(\mathbbm{1},\psi,\sigma)$ and fusion rules
\begin{equation}
\psi\times\psi=\mathbbm{1},\;\;
\psi\times\sigma=\sigma,\;\;
\sigma\times\sigma=\mathbbm{1}+\psi,
\end{equation}
where $\psi$ represents an electron and $\sigma$ carries a quasihole.
Because only $\sigma\times\sigma$ has multiple outcomes,
to enumerate $n$-quasihole states, we only need to
consider the fusion trees of $n$ $\sigma$ fields.
For example, for an even number of electrons, there are two degenerate
four-quasihole states,
\begin{equation}\label{eq:MR-4qh-tree}
|\Psi_a\rangle=\,
\begin{minipage}[c]{2cm}
\includegraphics[width=\linewidth]{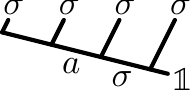}
\end{minipage},\hspace{1em}
a=\mathbbm{1}\text{ or }\psi.
\end{equation}
The structural similarity between the above fusion tree and the usual MPS
diagram~\cite{Schollwock11:DMRG-MPS} is not a coincidence,
but rather mandated by the radial ordering in the conformal 
correlator~\cite{Wu14:MPS}.
The $\mathbb{Z}_3$ RR CFT has a similar structure~\cite{Read99:RR,Ardonne07:RR},
with primary fields $(\mathbbm{1},\psi_1,\psi_2,\varepsilon,\sigma_1,\sigma_2)$.
Electrons and quasiholes are represented by $\psi_1$ and $\sigma_1$ 
fields~\cite{Wu14:Supplemental}, respectively.
There are again two four-quasihole states
\begin{equation}\label{eq:RR-4qh-tree}
|\Psi_a\rangle=\,
\begin{minipage}[c]{2cm}
\includegraphics[width=\linewidth]{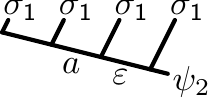}
\end{minipage},\hspace{1em}
a=\psi_1\text{ or }\sigma_2.
\end{equation}
In the following we examine the $\mathbb{Z}_{k\leq 3}$
RR states at fillings $\nu=\frac{k}{k+2}$,
and discuss the Gaffnian separately afterwards.

\begin{figure}[]
\centering
\includegraphics[]{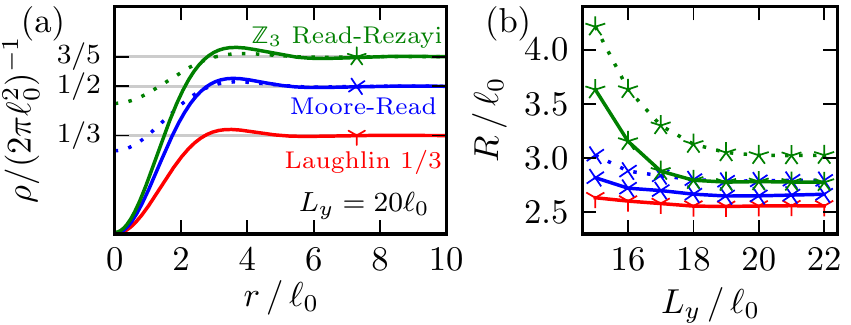}%
\caption{\label{fig:density-radius}
(a) Radial dependence of electron density $\rho$ near a quasihole, 
at $L_y=20\ell_0$.
For $\mathbb{Z}_{k=2,3}$ RR, we study both the $\frac{e}{k+2}$ 
(dotted) and the $\frac{ke}{k+2}$ quasiholes (solid curves).
(b) Quasihole radius $R$ as a function of cylinder perimeter $L_y$,
using the markers defined in (a).
}
\end{figure}

Before studying the braiding statistics, we first check whether the quasiholes 
are indeed localized excitations.
We check the charge $\nu e=\frac{ke}{k+2}$ Abelian quasiholes,
and for the $k>1$ theories, also the fundamental $\frac{e}{k+2}$ non-Abelian quasiholes.
In the electron density profile, we find a \emph{localized} and
\emph{isotropic} reduction 
around quasihole centers, which does not depend on the presence of other 
quasiholes sufficiently far away.
An example is given in Fig.~\ref{fig:density-heatmap} for
the RR $\frac{e}{5}$, $\frac{2e}{5}$ (two $\frac{e}{5}$ fused in $\sigma_2$), and 
$\frac{3e}{5}$ quasiholes.
The electron density displays small ripples in the periphery of a 
quasihole, as seen more clearly in the radial plots in 
Fig.~\ref{fig:density-radius}(a).
Following Ref.~\cite{Johri14:Quasihole},
we extract the quasihole radius $R$ from the second moment of the charge excess
distribution~\footnote{This definition of $R$ is very sensitive to the tail 
of the charge excess distribution. It clearly distinguishes quasihole sizes in 
different theories, even though the extent of the quasihole seems
comparable between different curves in Fig.~\ref{fig:density-radius}(a) by 
direct inspection.}.
This definition is susceptible to the thin-cylinder density 
wave background~\cite{Tao83:TT},
but it quickly converges as $L_y$
increases [Fig.~\ref{fig:density-radius}(b)].
We note that the quasihole size barely depends on $k$.
For $k>1$, the $\frac{e}{k+2}$ fundamental and the 
$\frac{ke}{k+2}$ Abelian quasiholes also have comparable sizes.
Listed in Table~\ref{tab:quasihole-data}, the numerical values are consistent 
with those previously reported for
$\mathbb{Z}_{k=1,2}$~\cite{Johri14:Quasihole,Prodan09:Braiding,Storni11:MR}.
For the $\mathbb{Z}_3$ RR quasiholes, our calculation provides the first 
radius estimate, $R\sim 3.0\,\ell_0$.
Strictly speaking this is a lower bound, as the quasiholes in the 
Coulomb ground state could be larger~\cite{Johri14:Quasihole}.

\begin{figure}[t]
\centering
\includegraphics[width=\linewidth]{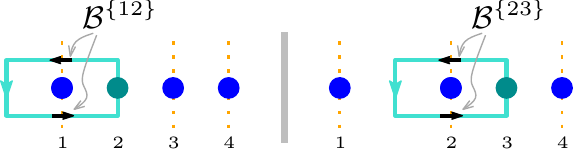}%
\caption{\label{fig:braiding}
Two braids of four-quasihole states, \{12\} and \{23\}.
We draw in yellow the branch cuts passing through each quasihole, and mark the 
half-braid $\mathcal{B}$ matrices by black arrows.}
\end{figure}

With the localized nature of the quasiholes established,
we now examine their braiding statistics.
For $\mathbb{Z}_{k=2,3}$ RR states, we consider the two braids $\{12\}$ and 
$\{23\}$ parametrized by the mobile quasihole $\eta$, as depicted in
Fig.~\ref{fig:braiding}.
We define the Berry connection over each $\mathrm{d}\eta$ 
segment along the braid
\begin{equation}
\mathcal{A}_{ab}(\eta;\mathrm{d}\eta)\equiv e^{-i \mathrm{d}\eta A_{ab}(\eta)}
\equiv\frac{\langle\Psi_a(\eta+\mathrm{d}\eta)|\Psi_b(\eta)\rangle}
{|\!|\Psi_a(\eta+\mathrm{d}\eta)|\!|\cdot |\!|\Psi_b(\eta)|\!|},
\end{equation}
in the unnormalized $\{|\Psi_a\rangle\}$ basis 
from Eqs.~(\ref{eq:MR-4qh-tree},~\ref{eq:RR-4qh-tree}).
Because of the nontrivial monodromy of the CFT correlator, $|\Psi_a(\eta)\rangle$ is 
multivalued in $\eta$. At each stationary quasihole, a branch cut runs 
vertically around the cylinder,
generating singularities in the Berry connection.
If we keep $\eta$ and $\eta+\mathrm{d}\eta$ on opposite sides of the cut while 
letting $\mathrm{d}\eta\rightarrow 0$ [Fig.~\ref{fig:braiding}],
the connection $\mathcal{A}(\eta;\mathrm{d}\eta)$ tends
to a constant matrix $\mathcal{B}$ \emph{not} equal to the identity:
\begin{align}\nonumber
\mathcal{B}_\text{MR}^{\{12\}}&=
\left[\begin{array}{cc}
1.0 & 0 \\
0 & 1.0\, i
\end{array}\right],\quad
\mathcal{B}_\text{RR}^{\{12\}}=
\left[\begin{array}{cc}
e^{0.6\pi i} & 0 \\
0 & 1.0
\end{array}\right],\\\label{eq:half-braid-numerical}
\mathcal{B}_\text{MR}^{\{23\}}&=
\left[\begin{array}{cc}
0.5005_{5}\!+\!0.5005_{5}i & 0.4994_{6}\!-\!0.4994_{6}i \\
0.4994_{6}\!-\!0.4994_{6}i & 0.5005_{5}\!+\!0.5005_{5}i
\end{array}\right],\\\nonumber
\mathcal{B}_\text{RR}^{\{23\}}&=
\left[\begin{array}{cc}
0.50008_{8}\!+\!0.36333_{6}i & 0.6357_{4}\!-\!0.4618_{3}i \\
0.63610_{9}\!-\!0.46215_{7}i & 0.19101_{2}\!+\!0.58786_{8}i
\end{array}\right]\!.
\end{align}
These matrices virtually coincide with the half-braid matrices~\cite{Moore89:CFT}
of the CFT correlators~\cite{Bonderson11:Plasma,Ardonne07:RR},
as in~\footnote{The RR matrices here are unitarily similar to 
Ref.~\cite{Ardonne07:RR}, due to their fusion tree choice
$\begin{minipage}[c]{0.7cm}
{\protect\includegraphics[width=\linewidth]{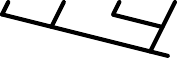}}
\end{minipage}$
versus our
$\begin{minipage}[c]{0.7cm}
{\protect\includegraphics[width=\linewidth]{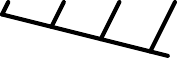}}
\end{minipage}$.}
\begin{gather}
\begin{aligned}
\begin{minipage}[c]{2cm}
\includegraphics[width=\linewidth]{fusion-tree-mr.pdf}
\end{minipage}
&=\left[\begin{array}{cc}
1 & 0 \\
0 & i
\end{array}\right]_{ba}
\begin{minipage}[c]{2cm}
\includegraphics[width=\linewidth]{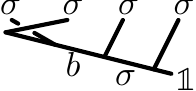}
\end{minipage}\\
&=\left[\begin{array}{cc}
\frac{1+i}{2} & \frac{1-i}{2} \\
\frac{1-i}{2} & \frac{1+i}{2}
\end{array}\right]_{ba}
\begin{minipage}[c]{2cm}
\includegraphics[width=\linewidth]{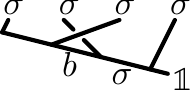}
\end{minipage},
\end{aligned}\\
\begin{aligned}
\begin{minipage}[c]{2cm}
\includegraphics[width=\linewidth]{fusion-tree-rr.pdf}
\end{minipage}
&=\left[\begin{array}{cc}
\omega^3 & 0 \\
0 & 1
\end{array}\right]_{ba}
\begin{minipage}[c]{2cm}
\includegraphics[width=\linewidth]{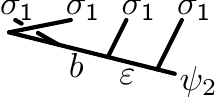}
\end{minipage}\\
&=\left[\begin{array}{cc}
\Phi\,\omega & \sqrt{\Phi}/\omega \\
\sqrt{\Phi}/\omega & \Phi\,\omega^2
\end{array}\right]_{ba}
\begin{minipage}[c]{2cm}
\includegraphics[width=\linewidth]{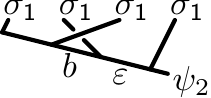}
\end{minipage},
\end{aligned}
\end{gather}
with $\omega=e^{i\pi/5}$ and $\Phi=\frac{\sqrt{5}-1}{2}$.
The subscripts in Eq.~\eqref{eq:half-braid-numerical} give the
last-digit deviations from the exact values~\footnote{In the actual calculations, we push the 
inert quasiholes outside of the braiding loop to infinity [see 
Eq.~\eqref{eq:fusion-abc}]. This essentially decouples the two conformal 
blocks for the $\{12\}$ braid, leading to very accurate (machine precision)
diagonal matrices for $\mathcal{B}^{\{12\}}$ at moderate $L_y$.}.

\begin{figure}[]
\centering
\includegraphics[]{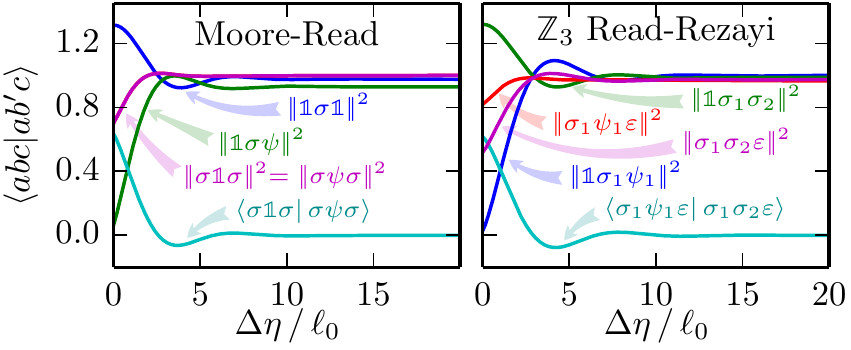}%
\caption{\label{fig:overlaps}
Dependence of the overlaps on the quasihole separation $\Delta\eta$ at $L_y=20\ell_0$.
$|\!|abc|\!|^2$ is shorthand for $\langle abc|abc\rangle$.
}
\end{figure}

The above half-braid $\mathcal{B}$ matrices, when squared, give the monodromy
prediction for the braiding statistics, namely, Ising and Fibonacci for 
$\mathbb{Z}_{k=2,3}$ RR, respectively.
These matrices should come out exactly as the CFT prediction, since we are 
implementing exactly the conformal blocks up to CFT truncation~\cite{Yurov90:TCSA}.
The actual braiding matrices are determined by the Wilson loop,
which, in addition to the $\mathcal{B}$ matrices, also depends on the Berry 
connections away from the branch cuts.
These nonsingular contributions are responsible for the potential discrepancy 
between monodromy and statistics, and are related to the overlap matrix 
$\langle\Psi_a|\Psi_b\rangle$ with fixed quasihole positions.
As detailed in Ref.~\cite{Bonderson11:Plasma}, \emph{if} at large quasihole 
separation $|\Delta\eta|$, the overlap converges exponentially fast to a 
constant diagonal matrix,
\begin{equation}\label{eq:screening}
\langle\Psi_a|\Psi_b\rangle=C_a\delta_{ab}+\mathcal{O}(e^{-|\Delta\eta|/\xi_{ab}}),
\text{ with }C_a\neq 0,
\end{equation}
then, except for the branch cuts, the Berry connection vanishes up to an 
exponentially small correction 
$\mathcal{A}_{ab}(\eta)\sim\mathcal{O}(e^{-|\Delta\eta|/\xi_{ab}})$
after subtracting the Aharonov-Bohm phase from the background magnetic field.
Hence, Eq.~\eqref{eq:screening} quantifies a sufficient condition for the 
equivalence between monodromy and statistics for well-separated quasiholes,
without the need to integrate $\mathcal{A}_{ab}(\eta)$ by brute force.
We now examine its validity for $\mathbb{Z}_{k=2,3}$ RR.
To simplify the functional form, we keep only two quasiholes at a finite 
separation $\Delta\eta$ and push others to infinity.
The resulting states are labeled by fusion trees
\begin{equation}\label{eq:fusion-abc}
|abc\rangle\equiv
\begin{minipage}[c]{1.35cm}
\includegraphics[width=\linewidth]{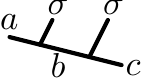}
\end{minipage}
\;\;\text{(for MR) or}\;\;
\begin{minipage}[c]{1.35cm}
\includegraphics[width=\linewidth]{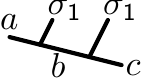}
\end{minipage}\text{ ($\mathbb{Z}_3$ RR)},
\end{equation}
and implemented by setting the MPS boundary conditions to the
leading eigenvectors of the MPS transfer matrix in topological
sectors $a$ and $c$~\cite{Estienne13:MPSLong}.
These states have the conformal-block normalization, up to a 
channel-independent overall constant.
Plotted in Fig.~\ref{fig:overlaps},
$\langle{abc}|{ab'c}\rangle$ indeed has 
the exponential convergence form of Eq.~\eqref{eq:screening} in all channels.
The correlation lengths can be estimated by curve fitting, or more 
conveniently, extracted from the spectral gaps of the transfer 
matrix~\cite{Wu14:Supplemental}.
Here we focus on the length scale $\xi_\text{ortho}$ associated with the 
decaying off-diagonal elements, characterizing the orthogonality between 
conformal blocks, and will report the diagonal ones 
elsewhere~\cite{Estienne14:Gaffnian}.
The numerical values are catalogued in Table~\ref{tab:quasihole-data}.
For MR, our results agree with Ref.~\cite{Baraban09:MR}.
For $\mathbb{Z}_3$ RR, combined with the $\mathcal{B}$ matrices shown 
earlier, the finite correlation lengths establish the quasiholes as Fibonacci 
anyons.
At the $\nu=12/5$ plateau\cite{Xia04:RR}, with a magnetic field of $5.4$ T,
$\xi\sim 3.4\,\ell_0$ translates to $0.038$ $\mu$m.
To put this in perspective, the inter-quasihole spacing in an interferometer
is on the order of $0.1$ $\mu$m~\cite{Rosenow12:Interferometer}
-- the quasiholes are not quite in the well-separated regime in this case.

\begin{figure}[]
\centering
\includegraphics[]{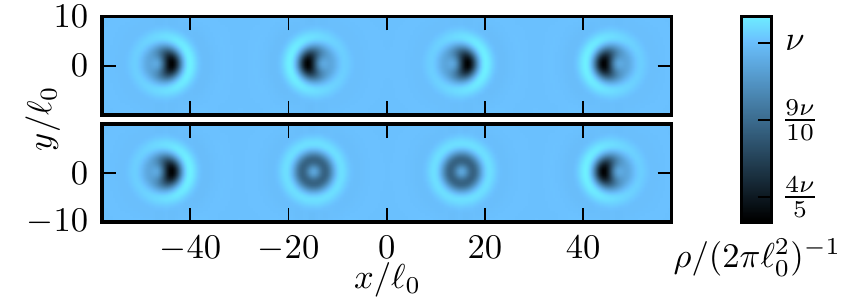}%
\caption{\label{fig:density-heatmap-gf}
Electron density for Gaffnian $\frac{e}{5}$ quasiholes in pair fusion channel 
$\mathbbm{1}$ (top) vs.\ $\varphi$ (bottom) [Eq.~\eqref{eq:Gf-4qh-tree}]
at $L_y=20\ell_0$.
}
\end{figure}

Finally, we go beyond the $\mathbb{Z}_k$ RR series, and study the 
Gaffnian~\cite{Simon07:Gaffnian} wave function at filling $\nu=2/5$.
It is derived from a nonunitary generalization of the Ising CFT, with 
primary fields $(\mathbbm{1},\psi,\sigma,\varphi)$~\cite{Wu14:Supplemental},
and is conjectured to describe a gapless state rather than a gapped 
topological phase~\cite{Read09:Adiabatic}.
There are two four-quasihole wave functions,
\begin{equation}\label{eq:Gf-4qh-tree}
|\Psi_a\rangle=\,
\begin{minipage}[c]{2cm}
\includegraphics[width=\linewidth]{fusion-tree-mr.pdf}
\end{minipage},\hspace{1em}
a=\mathbbm{1}\text{ or }\varphi.
\end{equation}
Each fundamental quasihole has charge $e/5$.
We start from the electron density profile of these states, as shown in 
Fig.~\ref{fig:density-heatmap-gf}.
In contrast to the $\mathbb{Z}_k$ RR states, the density profile here
exhibits a strong fusion-channel dependence, and also a dipole-like anisotropy for 
$|\Psi_{\mathbbm{1}}\rangle$, despite the clear separation of quasiholes.
Such local distinguishability of different conformal blocks persists even when 
the quasiholes are \emph{infinitely} separated.
This casts doubt on their topological degeneracy, although it remains 
unclear whether the density peculiarities are genuine, or artifacts at finite 
cylinder perimeter $L_y$~\cite{Wu14:Supplemental}.

\begin{figure}[]
\centering
\includegraphics[]{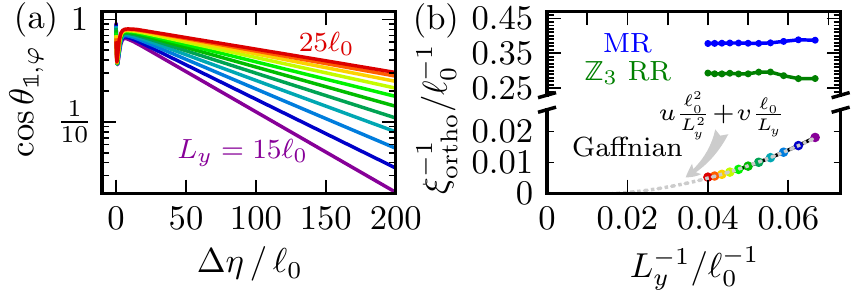}%
\caption{\label{fig:overlaps-gf}
(a) Asymptotic conformal-block orthogonality at large
quasihole separations for Gaffnian, color-coded by $L_y\in[15~..~25]\,\ell_0$.
(b) Comparison of the $L_y$ dependence of the associated length scales
between different theories. The Gaffnian curve has the best fit at $u=5.46(9)$, 
$v=-0.095(5)$.}
\end{figure}

Unlike the unitary $\mathbb{Z}_k$ theories, the conformal blocks for 
Gaffnian are not asymptotically orthogonal in general.
In place of the Berry connection, we consider the linear transform that 
relates $|\Psi_a(\eta)\rangle$ to $|\Psi_b(\eta+\mathrm{d}\eta)\rangle$,
\begin{equation}
\frac{|\Psi_b(\eta)\rangle}{|\!|\Psi_b(\eta)|\!|}
=\sum_a\widetilde{\mathcal{A}}_{ab}(\eta;\mathrm{d}\eta)
\frac{|\Psi_a(\eta+\mathrm{d}\eta)\rangle}{|\!|\Psi_a(\eta+\mathrm{d}\eta)|\!|},
\end{equation}
and we examine its behavior along the braids depicted in Fig.~\ref{fig:braiding}.
Across the branch cuts, the analogue of the unitary half-braid matrices $\mathcal{B}$
are $\widetilde{\mathcal{B}}\equiv\widetilde{\mathcal{A}}(\eta;\mathrm{d}\eta\rightarrow 0)$,
\begin{equation}
\begin{aligned}
\widetilde{\mathcal{B}}^{\{12\}}&=
\left[\begin{array}{cc}
e^{0.2\pi i} & 0 \\
0 & e^{0.4\pi i}
\end{array}\right],\\
\widetilde{\mathcal{B}}^{\{23\}}&=
\left[\begin{array}{cc}
1.616_2 & 0.746_1 + 1.028_1 i \\
0.746_1 + 1.028_1 i & -0.4995_5+1.537_1i
\end{array}\right]\!,
\end{aligned}
\end{equation}
in agreement with the CFT prediction
\begin{equation}
\begin{aligned}
\!\!
\begin{minipage}[c]{2cm}
\includegraphics[width=\linewidth]{fusion-tree-mr.pdf}
\end{minipage}
&=\left[\begin{array}{cc}
\omega & 0 \\
0 & \omega^2
\end{array}\right]_{ba}
\begin{minipage}[c]{2cm}
\includegraphics[width=\linewidth]{fusion-tree-mr-12.pdf}
\end{minipage}\\
&=\left[\begin{array}{cc}
1/\Phi & \omega^{\frac{3}{2}}/\Phi^{\frac{1}{2}} \\
\omega^{\frac{3}{2}}/\Phi^{\frac{1}{2}} & \omega^3/\Phi
\end{array}\right]_{ba}
\!\!\!\!
\begin{minipage}[c]{2cm}
\includegraphics[width=\linewidth]{fusion-tree-mr-23.pdf}
\end{minipage},
\end{aligned}
\end{equation}
with errors given in subscripts.
Again, this agreement indicates that our MPS correctly implements the 
conformal blocks.
The nonunitarity of $B^{\{23\}}$ comes from that of the
$F^{\sigma\sigma\sigma}_\sigma$ matrix~\cite{Ardonne11:YangLee}.
Away from the branch cuts, the behavior of $\widetilde{\mathcal{A}}$ is again 
controlled by the overlaps $\langle\Psi_a|\Psi_b\rangle$ as a function of
quasihole separations.
Ideally, we would like to examine the validity of Eq.~\eqref{eq:screening} for 
its individual matrix elements, as we do for the $\mathbb{Z}_k$ RR states.
In the $L_y\lesssim 25\ell_0$ regime accessible by MPS, however,
this calculation is 
plagued by finite-size effects, and we have trouble identifying its 
planar limit~\cite{Wu14:Supplemental}.
Fortunately, the conformal-block orthogonality measure
$\cos\theta_{\mathbbm{1},\varphi}\equiv
\frac{\langle{\sigma\mathbbm{1}\sigma}|{\sigma\varphi\sigma}\rangle}
{|\!|{\sigma\mathbbm{1}\sigma}|\!|\cdot|\!|{\sigma\varphi\sigma}|\!|}$
is immune from such artifacts.
Here we are keeping only two quasiholes at a finite separation
$\Delta\eta$ while the outer two are set to $\pm\infty$, and we use
the notation of Eq.~\eqref{eq:fusion-abc}.
As shown in Fig.~\ref{fig:overlaps-gf},
$\cos\theta_{\mathbbm{1},\varphi}$ decays exponentially
as $\Delta\eta$ increases,
but the associated length scale $\xi_\text{ortho}$ (as in
$\cos\theta_{\mathbbm{1},\varphi}\propto e^{-|\Delta\eta|/\xi_\text{ortho}}$)
diverges as $L_y\rightarrow\infty$,
in sharp contrast to the $\mathbb{Z}_k$ RR states.
Hence, in the planar limit, the conformal blocks in Eq.~\eqref{eq:Gf-4qh-tree} 
with an untwisted tree structure acquires orthogonality extremely slowly,
following a power law in $\Delta\eta$ (rather than exponentially),
signaling the breakdown of the screening condition Eq.~\eqref{eq:screening} 
for Gaffnian.
The power-law behavior is consistent with the conjectured 
gaplessness~\cite{Read09:Adiabatic}, and it largely 
rules out the possibility of defining a sensible braiding statistics for the 
Gaffnian quasiholes, since the nonuniversal corrections to the monodromy 
matrix are not exponentially small in quasihole separations.

To summarize, in this Letter we examined numerically the quasiholes in 
the $\mathbb{Z}_{k\leq 3}$ RR and the Gaffnian states using the MPS technique.
We provide, to our knowledge, the first size estimate for the $\mathbb{Z}_3$ RR quasiholes,
and also the first microscopic, quantitative verification of their Fibonacci 
nature.
We determine the correlation lengths associated with the exponential 
convergence of their braiding statistics.
In the context of topological quantum computing, these length scales
set the limit of the topological protection against decoherence in 
realistic systems~\cite{Baraban09:MR,Cheng09:Splitting,Bonderson09:Splitting}.
Our results also shed new light on the pathology of the Gaffnian wave function 
manifested in its quasiholes.

\emph{Acknowledgements}
We thank F.D.M.~Haldane, P.~Bonderson, S.H.~Simon, M.P.~Zaletel,
and S.~Johri for discussions.
BAB, NR, and YLW were supported by NSF CAREER DMR-095242, NSF-MRSEC DMR-0819860,
ONR-N00014-11-1-0635, MURI-130-6082, DARPA N66001-11-1-4110, Packard Foundation, and Keck grant.
NR was also supported by the Princeton Global Scholarship.
Numerical calculations were performed using the TIGRESS HPC facility at 
Princeton University.

\clearpage
\setcounter{equation}{0}
\setcounter{figure}{0}
\setcounter{table}{0}
\makeatletter

\renewcommand{\theequation}{S\arabic{equation}}
\renewcommand{\thefigure}{S\arabic{figure}}
\renewcommand{\bibnumfmt}[1]{[S#1]}
\renewcommand{\citenumfont}[1]{S#1}

\section{Supplemental Material}

Here we address some technical aspects of the matrix product states 
(MPS) for non-Abelian quasiholes derived from conformal field theory (CFT) 
correlators.
Similar to the treatment in the main text, we will leave the compactified U(1) 
boson implicit in the CFT description.

\subsection{Fusion rules for the $\mathbb{Z}_3$ Read-Rezayi state}
The $\mathbb{Z}_3$ Read-Rezayi state can be described by the $\mathbb{Z}_3$ 
parafermion conformal field theory~\cite{Read99:RR:Supplemental},
also known as the minimal model $\mathcal{M}(5,6)$~\cite{DiFrancesco99:Yellow:Supplemental},
with central charge $c=\frac{4}{5}$.
The primary fields of this CFT are 
$(\mathbbm{1},\psi_1,\psi_2,\varepsilon,\sigma_1,\sigma_2)$,
with scaling dimensions 
$(0,\frac{2}{3},\frac{2}{3},\frac{2}{5},\frac{1}{15},\frac{1}{15})$.
The $\psi_1$ (resp.\ $\sigma_1$) field represents an electron (resp.\ a 
quasihole).
The fusion rules of these fields are
\begin{equation*}
\begin{minipage}[h]{0.9\linewidth}
\begin{ruledtabular}
\begin{tabular}{c|cccccc}
 & $\mathbbm{1}$ & $\psi_1$ & $\psi_2$
& $\varepsilon$ & $\sigma_1$ & $\sigma_2$ \\\hline\hline
$\psi_1$ & $\psi_1$ & $\psi_2$ & $\mathbbm{1}$
& $\sigma_2$ & $\varepsilon$ & $\sigma_1$ \\\hline
$\sigma_1$ & $\sigma_1$ & $\varepsilon$ & $\sigma_2$
& $\psi_2+\sigma_1$ & $\psi_1+\sigma_2$ & $\mathbbm{1}+\varepsilon$
\end{tabular}
\end{ruledtabular}
\end{minipage}
\end{equation*}

\subsection{Fusion rules for the Gaffnian state}
The Gaffnian wave function is described~\cite{Simon07:Gaffnian:Supplemental} by
the non-unitary CFT minimal model $\mathcal{M}(3,5)$,
with central charge $c=-\frac{3}{5}$.
The primary fields of this CFT are $(\mathbbm{1},\psi,\sigma,\varphi)$,
with scaling dimensions $(0,\frac{3}{4},-\frac{1}{20},\frac{1}{5})$.
The $\psi$ (resp.\ $\sigma$) field represents an electron (resp.\ a quasihole),
with fusion rules
\begin{equation*}
\begin{minipage}[h]{0.6\linewidth}
\begin{ruledtabular}
\begin{tabular}{c|cccc}
 & $\mathbbm{1}$ & $\psi$ & $\sigma$ & $\varphi$ \\\hline\hline
$\psi$ & $\psi$ & $\mathbbm{1}$ & $\varphi$ & $\sigma$ \\\hline
$\sigma$ & $\sigma$ & $\varphi$ & $\mathbbm{1}+\varphi$ & $\psi+\sigma$
\end{tabular}
\end{ruledtabular}
\end{minipage}
\end{equation*}

\subsection{The MPS transfer matrix}
Following the notation of Ref.~\cite{Estienne13:MPSLong:Supplemental}, we consider the 
transfer matrix $E=\sum_m^{0,1}(B^m)^*\otimes B^m$, where the $B^m$ matrix
is associated with an empty $(m=0)$ or occupied $(m=1)$ Landau orbital in the 
MPS.
The transfer matrix is the basic building block of any generic wave function 
overlap $\langle\Psi|\Psi'\rangle$.
It acts on a direct product of two copies of the truncated conformal Hilbert 
space, one copy for $\langle\Psi|$, and the other for $|\Psi'\rangle$.
From the fusion rules, we find that the CFT Hilbert space can be naturally 
split into two sectors, each being closed under fusion with the electron 
(although they are connected by fusion with the quasihole).
We refer to them as the ``vac'' and the ``qh'' sectors:
\begin{equation*}
\begin{minipage}[h]{0.8\linewidth}
\begin{ruledtabular}
\begin{tabular}{c|cc}
 & vac & qh \\\hline\hline
Moore-Read & $\mathbbm{1},\psi$ & $\sigma$ \\\hline
$\mathbb{Z}_3$ Read-Rezayi & $\mathbbm{1},\psi_1,\psi_2$ & $\varepsilon,\sigma_1,\sigma_2$ \\\hline
Gaffnian & $\mathbbm{1},\psi$ & $\sigma,\varphi$
\end{tabular}
\end{ruledtabular}
\end{minipage}
\end{equation*}
The $B^m$ matrices are block-diagonal in the sector index,
$B^m=\bigoplus_\alpha B^m_\alpha$, with $\alpha$ summed over \{vac, qh\}.
Therefore, the transfer matrix is also block-diagonal,
\begin{equation}\label{eq:transfer-matrix}
E=\bigoplus_{\alpha,\beta}E_{\alpha,\beta},\text{ with }
E_{\alpha,\beta}=\sum_m(B^m_\alpha)^*\otimes B^m_\beta.
\end{equation}
We denote by $\lambda^{(i)}_{\alpha,\beta}$ the $i$-th largest eigenvalue of 
$E_{\alpha,\beta}$.
The MPS auxiliary space is constructed from the truncated conformal Hilbert 
space, and the truncation is constrained by the entanglement area law.
In our calculations, we have to deal with transfer matrix blocks 
(after various reductions~\cite{Estienne13:MPSLong:Supplemental})
with dimensions as large as
\begin{equation*}
\begin{minipage}[h]{\linewidth}
\begin{ruledtabular}
\begin{tabular}{c|ccc} 
 & (vac, vac) & (vac, qh) & (qh, qh)\\\hline\hline
Moore-Read & $1.1\times 10^7$ & $1.5\times 10^7$ & $2.0\times 10^7$ \\\hline
$\mathbb{Z}_3$ Read-Rezayi & $3.6\times 10^7$ & $5.5\times 10^7$ & $8.4\times 10^7$ \\\hline
Gaffnian & $1.3\times 10^7$ & $2.0\times 10^7$ & $3.0\times 10^7$
\end{tabular}
\end{ruledtabular}
\end{minipage}
\end{equation*}
Incidentally, for the braiding and the overlap calculations, we have to work
on the full direct product space without symmetry reduction, the dimension of
which can be up to $25$ times as large as the sizes mentioned in the previous 
table.

\subsection{Overlap calculation}

As explained in the main text, the central object in our braiding study is the 
overlap matrix $\langle abc|ab'c\rangle$, and we are particularly interested 
in its exponential convergence
\begin{equation}\label{eq:screening-abc}
\langle abc|ab'c\rangle=C_{abc}\delta_{bb'}
+\mathcal{O}(e^{-|\Delta\eta|/\xi_{\langle abc|ab'c\rangle}}).
\end{equation}
In the following we outline the calculation of the overlap matrix using the 
MPS technique, and also discuss the determination of the correlation lengths.
Recall from the main text that the state
\begin{equation}\label{eq:fusion-abc}
|abc\rangle\equiv
\begin{minipage}[c]{1.35cm}
\includegraphics[width=\linewidth]{fusion-tree-mr-abc.pdf}
\end{minipage}
\end{equation}
involves two localized quasiholes at a finite separation $\Delta\eta$, and the 
topological charges $a$ and $c$ represent extra quasiholes pushed to the ends 
of the infinite cylinder.
Diagrammatically, the MPS for $|abc\rangle$ is given by~\cite{Zaletel12:MPS:Supplemental,Estienne13:MPSLong:Supplemental}
\begin{equation*}
\includegraphics[width=0.9\linewidth]{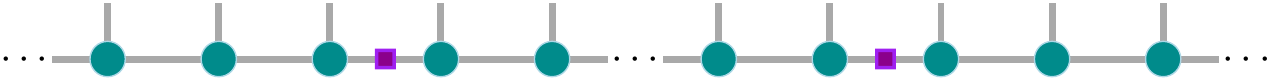}.
\end{equation*}
Here, the orbital $B^m$ matrices are represented by the green circles, with 
the occupation number $m=0,1$ carried by the upward-pointing leg,
and the quasihole matrices are represented by the purple squares.
Each quasihole matrix depends on both the quasihole position and the fusion 
channel context, i.e. the topological charges before and after the $\sigma$ 
field insertion in the fusion tree,
and it is inserted into the matrix product at the correct time-ordered 
positions~\cite{Zaletel12:MPS:Supplemental}.
Technical details of the construction of the quasihole matrix will be 
addressed in a forthcoming paper~\cite{Wu14:MPS:Supplemental}.
The overlap $\langle abc|ab'c\rangle$ is computed by contracting
\begin{equation*}
\includegraphics[width=0.9\linewidth]{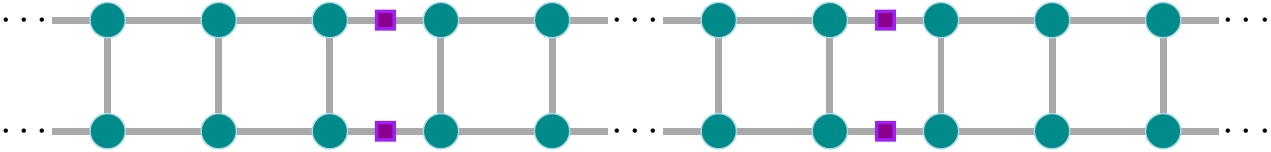}.
\end{equation*}
Here, the upper (lower) chain represents the $\langle abc|$ ($|ab'c\rangle$) 
state, respectively, and in the ladder-like structure, each rung corresponds 
to the transfer matrix $E$ over a single orbital [Eq.~\eqref{eq:transfer-matrix}].
Although not marked explicitly in the above diagrams, the fusion channel 
dependence enters through the quasihole insertions (purple squares) as well as 
the boundary conditions.
The contraction of the above tensor network can be significantly simplified on 
an infinite cylinder, as detailed in Ref.~\onlinecite{Estienne13:MPSLong:Supplemental}.
Essentially, an \emph{infinitely} repeated action of the transfer matrix can be 
accurately represented by its projection into the subspace of its largest 
eigenvalue in the relevant sector.
For the overlap $\langle abc|ab'c\rangle$ with a finite quasihole separation 
$\Delta\eta$, as shown in Fig.~4 of the main text, this simplification applies 
only to the peripheral regions outside of the two quasihole insertions.
Between the two quasiholes, we have to contract the transfer matrices by brute 
force.
However, in the limit of large $\Delta\eta$, asymptotically the overlap is 
still controlled by the leading few eigenmodes of the transfer matrix,
and the associated correlation lengths can be simply determined from the 
spectral gaps of the transfer matrix, without resorting to curve fitting.

We now explain this using three representative examples.
First, consider the off-diagonal element
$\langle\sigma_1\psi_1\varepsilon|\sigma_1\sigma_2\varepsilon\rangle$ for the 
$\mathbb{Z}_3$ Read-Rezayi state.
In this case, the action of the transfer matrix over the $\Delta\eta$ interval 
is confined to the (vac, qh) sector of the product space,
while its action outside of the $\Delta\eta$ interval is purely in the 
(qh, qh) sector.
At large $\Delta\eta$, we must have
\begin{equation}
\langle\sigma_1\psi_1\varepsilon|\sigma_1\sigma_2\varepsilon\rangle\sim
\left(\frac{\lambda^{(1)}_\text{vac,qh}}{\lambda^{(1)}_\text{qh,qh}}\right)^{\Delta\eta/\gamma}.
\end{equation}
Here $\gamma=2\pi\ell_0^2/L_y$ is the separation between adjacent Landau 
orbitals, while $\lambda^{(1)}_\text{vac,qh}$ and $\lambda^{(1)}_\text{qh,qh}$ 
are the largest eigenvalues of the transfer matrix in sectors (vac,qh) and 
(qh, qh), resp.
The correlation length is then given by
\begin{equation}
\xi_\text{ortho}
=\left[\frac{L_y}{2\pi\ell_0^2}
\log\left(\frac{\lambda^{(1)}_\text{qh,qh}}{\lambda^{(1)}_\text{vac,qh}}
\right)\right]^{-1}.
\end{equation}
As the second example, we consider the norm $|\!|\mathbbm{1}\sigma\psi|\!|^2$
for the Moore-Read state.
To the leading order, we have
\begin{equation}\label{eq:mr-overlap-120}
|\!|\mathbbm{1}\sigma\psi|\!|^2\sim
\left(\frac{\lambda^{(1)}_\text{qh,qh}}{\lambda^{(1)}_\text{vac,vac}}\right)^{\Delta\eta/\gamma},
\end{equation}
while $\lambda^{(1)}_\text{vac,vac}$ and $\lambda^{(1)}_\text{qh,qh}$ 
are the largest eigenvalues of the transfer matrix in sectors (vac,vac) and 
(qh, qh), resp.
For $|\!|\mathbbm{1}\sigma\psi|\!|^2$ to approach a non-zero constant value 
when $\Delta\eta\rightarrow\infty$ as in Eq.~\eqref{eq:screening-abc}, we need 
to have $\lambda^{(1)}_\text{vac,vac}=\lambda^{(1)}_\text{qh,qh}$.
This turns out to be true for the $\mathbb{Z}_{k=2,3}$ Read-Rezayi states, 
up to small finite-size corrections (see below).
To characterize the exponential convergence of the norm,
we have to consider the second largest 
eigenvalue in the (qh,qh) channel, $\lambda^{(2)}_\text{qh,qh}$.
The associated correlation length is given by
\begin{equation}\label{eq:xi-120}
\xi_\text{qh}=\left[\frac{L_y}{2\pi\ell_0^2}
\log\left(\frac{\lambda^{(1)}_\text{qh,qh}}{\lambda^{(2)}_\text{qh,qh}}
\right)\right]^{-1}.
\end{equation}
Finally, the correlation length associated with the norm 
$|\!|\sigma\psi\sigma|\!|^2$ for the Moore-Read state is similarly given by
\begin{equation}\label{eq:xi-212}
\xi_\text{vac}=\left[\frac{L_y}{2\pi\ell_0^2}
\log\left(\frac{\lambda^{(1)}_\text{vac,vac}}{\lambda^{(2)}_\text{vac,vac}}
\right)\right]^{-1}.
\end{equation}
The numerical calculations of $\xi_\text{vac}$ and $\xi_\text{qh}$ are more 
challenging than $\xi_\text{ortho}$, since they depend on subleading 
eigenvalues of the transfer matrix.
A detailed numerical study will be reported in a future 
paper~\cite{Estienne14:Gaffnian:Supplemental}.

\begin{figure}[]
\centering
\includegraphics[]{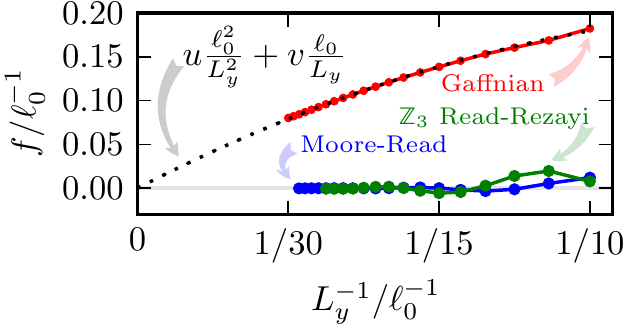}%
\caption{\label{fig:anticonfinement}
Asymptotic repulsion $f$ between two plasma charges representing 
pinned quasiholes. The Gaffnian curve is fitted by the zero-intercept 
quadratic formula $f \ell_0=u\frac{\ell_0^2}{L_y^2}+v\frac{\ell_0}{L_y}$.
The best fit has $u=-8.6(2)$ and $v=2.66(1)$, with the standard error in the 
last digit given in parentheses.}
\end{figure}

We now examine the $\lambda^{(1)}_\text{vac,vac}=\lambda^{(1)}_\text{qh,qh}$ 
condition more carefully.
To have a physical understanding of its implication, we adopt the plasma 
analogy and reinterpret the overlap in Eq.~\eqref{eq:mr-overlap-120} as the 
partition function $e^{-F(\Delta\eta)}$ with two pinned charges 
representing the two quasiholes at a separation $\Delta\eta$.
The derivative of the free energy $F(\Delta\eta)$ with respect to $\Delta\eta$ 
gives an effective force between the plasma charges
\begin{equation}
f=-\frac{\mathrm{d}F}{\mathrm{d}\Delta\eta}\sim 
\frac{L_y}{2\pi\ell_0^2}
\log\left(\frac{\lambda^{(1)}_\text{qh,qh}}{\lambda^{(1)}_\text{vac,vac}}\right).
\end{equation}
Therefore, if $\lambda^{(1)}_\text{vac,vac}\neq\lambda^{(1)}_\text{qh,qh}$,
the two plasma charges representing quasiholes are subject to an 
asymptotically constant confining (if $f<0$) or anti-confining (if $f>0$) 
force that persists even in the limit of infinite separation.
The numerical data are shown in Fig.~\ref{fig:anticonfinement}.
For the Moore-Read and the $\mathbb{Z}_3$ Read-Rezayi states,
$\lambda^{(1)}_\text{vac,vac}$ and $\lambda^{(1)}_\text{qh,qh}$ quickly 
converge as $L_y$ increases.
In contrast, the Gaffnian state features an asymptotic repulsion between 
infinitely separated plasma charges at a finite cylinder perimeter $L_y$,
although it seems to die off in the planar limit $L_y\rightarrow\infty$.
This makes it very hard to extract a meaningful correlation length for the 
diagonal elements of the overlap matrix similar to Eq.~\eqref{eq:xi-120}.
Fortunately, we can still analyze the correlation length associated with the 
asymptotic orthogonality of conformal blocks, as discussed in the main text.

\subsection{Electron density profile around Gaffnian quasiholes}

\begin{figure}[]
\centering
\includegraphics[]{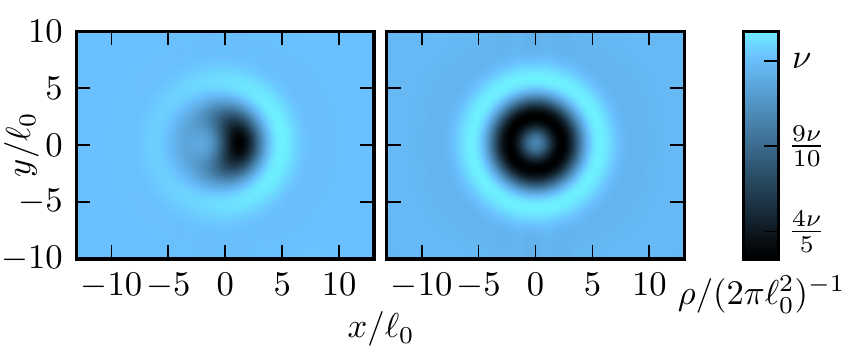}%
\caption{\label{fig:density-heatmap-gf-single}
Electron density around a single Gaffnian $\frac{e}{5}$ quasihole in 
the $|\mathbbm{1}\sigma\rangle$ (left panel) and the $|\sigma\varphi\rangle$ 
(right panel) channels, on an infinitely long cylinder with perimeter 
$L_y=20\ell_0$.}
\end{figure}

Here we show more details of the peculiarities in the electron density profile 
of the Gaffnian quasiholes.
As noted in the main text, conformal blocks in different fusion channels
are locally distinguishable despite the clear separation between quasiholes.
This effect persists even when we push the quasihole separations to infinity,
leaving only a single fully isolated quasihole.
In this limit, the conformal blocks can be labeled by fusion tree segments
\begin{equation}
|ab\rangle\equiv
\begin{minipage}[c]{1.0cm}
\includegraphics[width=\linewidth]{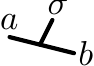}.
\end{minipage}
\end{equation}
We only need to consider $|ab\rangle=|\mathbbm{1}\sigma\rangle$
and $|\sigma\varphi\rangle$, since all the other 
possibilities can be obtained by either fusing (trivially) with $\psi$,
or flipping the cylinder axis $x\rightarrow -x$.
Fig.~\ref{fig:density-heatmap-gf-single} shows the electron density profile for each case.
The anisotropic dipole structure is clearly visible for 
$|\mathbbm{1}\sigma\rangle$, in contrast to the isotropic 
$|\sigma\varphi\rangle$.

\begin{figure}[]
\centering
\includegraphics[]{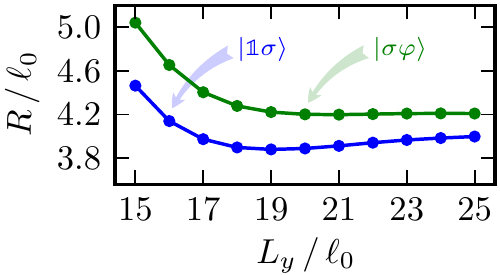}%
\caption{\label{fig:radius-gf}
Radii of Gaffnian quasiholes in $|\mathbbm{1}\sigma\rangle$ and 
$|\sigma\varphi\rangle$ channels, as a function of the cylinder perimeter $L_y$.}
\end{figure}

Similar to the $\mathbb{Z}_{k\leq 3}$ Read-Rezayi quasiholes analyzed in the 
main text, we estimate the quasihole radius from the second moment of the 
charge excess distribution [Fig.~\ref{fig:radius-gf}].
We have also examined the 
Abelian charge $\frac{2e}{5}$ quasihole obtained by fusing two $\frac{e}{5}$ 
quasiholes in the $\mathbbm{1}$ channel. We find a localized and isotropic 
density reduction around each Abelian quasihole, but this calculation turns 
out to be rather susceptible to the conformal Hilbert space truncation, and we 
have trouble reaching convergence in the radius calculation.
As a final comment, we note that the peculiarities observed in the density 
profile are likely related to the leading eigenvalue mismatch discussed in the 
previous section, and are possibly artifacts at finite cylinder perimeter $L_y$.
Unfortunately, we cannot resolve this issue using the current MPS approach,
due to the fundamental constraint on $L_y$ from the area law of quantum 
entanglement.


\begin{thebibliography}{69}%
\makeatletter
\providecommand \@ifxundefined [1]{%
 \@ifx{#1\undefined}
}%
\providecommand \@ifnum [1]{%
 \ifnum #1\expandafter \@firstoftwo
 \else \expandafter \@secondoftwo
 \fi
}%
\providecommand \@ifx [1]{%
 \ifx #1\expandafter \@firstoftwo
 \else \expandafter \@secondoftwo
 \fi
}%
\providecommand \natexlab [1]{#1}%
\providecommand \enquote  [1]{``#1''}%
\providecommand \bibnamefont  [1]{#1}%
\providecommand \bibfnamefont [1]{#1}%
\providecommand \citenamefont [1]{#1}%
\providecommand \href@noop [0]{\@secondoftwo}%
\providecommand \href [0]{\begingroup \@sanitize@url \@href}%
\providecommand \@href[1]{\@@startlink{#1}\@@href}%
\providecommand \@@href[1]{\endgroup#1\@@endlink}%
\providecommand \@sanitize@url [0]{\catcode `\\12\catcode `\$12\catcode
  `\&12\catcode `\#12\catcode `\^12\catcode `\_12\catcode `\%12\relax}%
\providecommand \@@startlink[1]{}%
\providecommand \@@endlink[0]{}%
\providecommand \url  [0]{\begingroup\@sanitize@url \@url }%
\providecommand \@url [1]{\endgroup\@href {#1}{\urlprefix }}%
\providecommand \urlprefix  [0]{URL }%
\providecommand \Eprint [0]{\href }%
\providecommand \doibase [0]{http://dx.doi.org/}%
\providecommand \selectlanguage [0]{\@gobble}%
\providecommand \bibinfo  [0]{\@secondoftwo}%
\providecommand \bibfield  [0]{\@secondoftwo}%
\providecommand \translation [1]{[#1]}%
\providecommand \BibitemOpen [0]{}%
\providecommand \bibitemStop [0]{}%
\providecommand \bibitemNoStop [0]{.\EOS\space}%
\providecommand \EOS [0]{\spacefactor3000\relax}%
\providecommand \BibitemShut  [1]{\csname bibitem#1\endcsname}%
\let\auto@bib@innerbib\@empty
\bibitem [{\citenamefont {Kitaev}(2003)}]{Kitaev03:TQC}%
  \BibitemOpen
  \bibfield  {author} {\bibinfo {author} {\bibfnamefont {A.~Y.}\ \bibnamefont
  {Kitaev}},\ }\href {\doibase 10.1016/S0003-4916(02)00018-0} {\bibfield
  {journal} {\bibinfo  {journal} {Annals of Physics}\ }\textbf {\bibinfo
  {volume} {303}},\ \bibinfo {pages} {2} (\bibinfo {year} {2003})}\BibitemShut
  {NoStop}%
\bibitem [{\citenamefont {Freedman}\ \emph
  {et~al.}(2002{\natexlab{a}})\citenamefont {Freedman}, \citenamefont
  {Larsen},\ and\ \citenamefont {Wang}}]{Freedman02:Universal1}%
  \BibitemOpen
  \bibfield  {author} {\bibinfo {author} {\bibfnamefont {M.~H.}\ \bibnamefont
  {Freedman}}, \bibinfo {author} {\bibfnamefont {M.}~\bibnamefont {Larsen}}, \
  and\ \bibinfo {author} {\bibfnamefont {Z.}~\bibnamefont {Wang}},\ }\href
  {\doibase 10.1007/s002200200645} {\bibfield  {journal} {\bibinfo  {journal}
  {Communications in Mathematical Physics}\ }\textbf {\bibinfo {volume}
  {227}},\ \bibinfo {pages} {605} (\bibinfo {year}
  {2002}{\natexlab{a}})}\BibitemShut {NoStop}%
\bibitem [{\citenamefont {Freedman}\ \emph
  {et~al.}(2002{\natexlab{b}})\citenamefont {Freedman}, \citenamefont
  {Larsen},\ and\ \citenamefont {Wang}}]{Freedman02:Universal2}%
  \BibitemOpen
  \bibfield  {author} {\bibinfo {author} {\bibfnamefont {M.~H.}\ \bibnamefont
  {Freedman}}, \bibinfo {author} {\bibfnamefont {M.~J.}\ \bibnamefont
  {Larsen}}, \ and\ \bibinfo {author} {\bibfnamefont {Z.}~\bibnamefont
  {Wang}},\ }\href {\doibase 10.1007/s002200200636} {\bibfield  {journal}
  {\bibinfo  {journal} {Communications in Mathematical Physics}\ }\textbf
  {\bibinfo {volume} {228}},\ \bibinfo {pages} {177} (\bibinfo {year}
  {2002}{\natexlab{b}})}\BibitemShut {NoStop}%
\bibitem [{\citenamefont {Hormozi}\ \emph {et~al.}(2007)\citenamefont
  {Hormozi}, \citenamefont {Zikos}, \citenamefont {Bonesteel},\ and\
  \citenamefont {Simon}}]{Hormozi07:Compiling}%
  \BibitemOpen
  \bibfield  {author} {\bibinfo {author} {\bibfnamefont {L.}~\bibnamefont
  {Hormozi}}, \bibinfo {author} {\bibfnamefont {G.}~\bibnamefont {Zikos}},
  \bibinfo {author} {\bibfnamefont {N.~E.}\ \bibnamefont {Bonesteel}}, \ and\
  \bibinfo {author} {\bibfnamefont {S.~H.}\ \bibnamefont {Simon}},\ }\href
  {\doibase 10.1103/PhysRevB.75.165310} {\bibfield  {journal} {\bibinfo
  {journal} {Phys. Rev. B}\ }\textbf {\bibinfo {volume} {75}},\ \bibinfo
  {pages} {165310} (\bibinfo {year} {2007})}\BibitemShut {NoStop}%
\bibitem [{\citenamefont {Nayak}\ \emph {et~al.}(2008)\citenamefont {Nayak},
  \citenamefont {Stern}, \citenamefont {Freedman},\ and\ \citenamefont
  {Das~Sarma}}]{Nayak08:RMP}%
  \BibitemOpen
  \bibfield  {author} {\bibinfo {author} {\bibfnamefont {C.}~\bibnamefont
  {Nayak}}, \bibinfo {author} {\bibfnamefont {A.}~\bibnamefont {Stern}},
  \bibinfo {author} {\bibfnamefont {M.}~\bibnamefont {Freedman}}, \ and\
  \bibinfo {author} {\bibfnamefont {S.}~\bibnamefont {Das~Sarma}},\ }\href
  {\doibase 10.1103/RevModPhys.80.1083} {\bibfield  {journal} {\bibinfo
  {journal} {Reviews of Modern Physics}\ }\textbf {\bibinfo {volume} {80}},\
  \bibinfo {pages} {1083} (\bibinfo {year} {2008})}\BibitemShut {NoStop}%
\bibitem [{\citenamefont {Levin}\ and\ \citenamefont
  {Wen}(2005)}]{Levin05:StringNet}%
  \BibitemOpen
  \bibfield  {author} {\bibinfo {author} {\bibfnamefont {M.~A.}\ \bibnamefont
  {Levin}}\ and\ \bibinfo {author} {\bibfnamefont {X.-G.}\ \bibnamefont
  {Wen}},\ }\href {\doibase 10.1103/PhysRevB.71.045110} {\bibfield  {journal}
  {\bibinfo  {journal} {Phys. Rev. B}\ }\textbf {\bibinfo {volume} {71}},\
  \bibinfo {pages} {045110} (\bibinfo {year} {2005})}\BibitemShut {NoStop}%
\bibitem [{\citenamefont {Fu}\ and\ \citenamefont
  {Kane}(2008)}]{Fu08:Proximity}%
  \BibitemOpen
  \bibfield  {author} {\bibinfo {author} {\bibfnamefont {L.}~\bibnamefont
  {Fu}}\ and\ \bibinfo {author} {\bibfnamefont {C.~L.}\ \bibnamefont {Kane}},\
  }\href {\doibase 10.1103/PhysRevLett.100.096407} {\bibfield  {journal}
  {\bibinfo  {journal} {Phys. Rev. Lett.}\ }\textbf {\bibinfo {volume} {100}},\
  \bibinfo {pages} {096407} (\bibinfo {year} {2008})}\BibitemShut {NoStop}%
\bibitem [{\citenamefont {Lutchyn}\ \emph {et~al.}(2010)\citenamefont
  {Lutchyn}, \citenamefont {Sau},\ and\ \citenamefont
  {Das~Sarma}}]{Lutchyn10:Majorana}%
  \BibitemOpen
  \bibfield  {author} {\bibinfo {author} {\bibfnamefont {R.~M.}\ \bibnamefont
  {Lutchyn}}, \bibinfo {author} {\bibfnamefont {J.~D.}\ \bibnamefont {Sau}}, \
  and\ \bibinfo {author} {\bibfnamefont {S.}~\bibnamefont {Das~Sarma}},\ }\href
  {\doibase 10.1103/PhysRevLett.105.077001} {\bibfield  {journal} {\bibinfo
  {journal} {Phys. Rev. Lett.}\ }\textbf {\bibinfo {volume} {105}},\ \bibinfo
  {pages} {077001} (\bibinfo {year} {2010})}\BibitemShut {NoStop}%
\bibitem [{\citenamefont {Oreg}\ \emph {et~al.}(2010)\citenamefont {Oreg},
  \citenamefont {Refael},\ and\ \citenamefont {von Oppen}}]{Oreg10:Majorana}%
  \BibitemOpen
  \bibfield  {author} {\bibinfo {author} {\bibfnamefont {Y.}~\bibnamefont
  {Oreg}}, \bibinfo {author} {\bibfnamefont {G.}~\bibnamefont {Refael}}, \ and\
  \bibinfo {author} {\bibfnamefont {F.}~\bibnamefont {von Oppen}},\ }\href
  {\doibase 10.1103/PhysRevLett.105.177002} {\bibfield  {journal} {\bibinfo
  {journal} {Phys. Rev. Lett.}\ }\textbf {\bibinfo {volume} {105}},\ \bibinfo
  {pages} {177002} (\bibinfo {year} {2010})}\BibitemShut {NoStop}%
\bibitem [{\citenamefont {Alicea}\ \emph {et~al.}(2011)\citenamefont {Alicea},
  \citenamefont {Oreg}, \citenamefont {Refael}, \citenamefont {von Oppen},\
  and\ \citenamefont {Fisher}}]{Alicea11:Network}%
  \BibitemOpen
  \bibfield  {author} {\bibinfo {author} {\bibfnamefont {J.}~\bibnamefont
  {Alicea}}, \bibinfo {author} {\bibfnamefont {Y.}~\bibnamefont {Oreg}},
  \bibinfo {author} {\bibfnamefont {G.}~\bibnamefont {Refael}}, \bibinfo
  {author} {\bibfnamefont {F.}~\bibnamefont {von Oppen}}, \ and\ \bibinfo
  {author} {\bibfnamefont {M.~P.~A.}\ \bibnamefont {Fisher}},\ }\href {\doibase
  10.1038/NPHYS1915} {\bibfield  {journal} {\bibinfo  {journal} {Nature
  Physics}\ }\textbf {\bibinfo {volume} {7}},\ \bibinfo {pages} {412} (\bibinfo
  {year} {2011})}\BibitemShut {NoStop}%
\bibitem [{\citenamefont {Alicea}(2012)}]{Alicea12:Majorana}%
  \BibitemOpen
  \bibfield  {author} {\bibinfo {author} {\bibfnamefont {J.}~\bibnamefont
  {Alicea}},\ }\href {\doibase 10.1088/0034-4885/75/7/076501} {\bibfield
  {journal} {\bibinfo  {journal} {Reports on Progress in Physics}\ }\textbf
  {\bibinfo {volume} {75}},\ \bibinfo {pages} {076501} (\bibinfo {year}
  {2012})}\BibitemShut {NoStop}%
\bibitem [{\citenamefont {Mourik}\ \emph {et~al.}(2012)\citenamefont {Mourik},
  \citenamefont {Zuo}, \citenamefont {Frolov}, \citenamefont {Plissard},
  \citenamefont {Bakkers},\ and\ \citenamefont
  {Kouwenhoven}}]{Mourik12:Majorana}%
  \BibitemOpen
  \bibfield  {author} {\bibinfo {author} {\bibfnamefont {V.}~\bibnamefont
  {Mourik}}, \bibinfo {author} {\bibfnamefont {K.}~\bibnamefont {Zuo}},
  \bibinfo {author} {\bibfnamefont {S.~M.}\ \bibnamefont {Frolov}}, \bibinfo
  {author} {\bibfnamefont {S.~R.}\ \bibnamefont {Plissard}}, \bibinfo {author}
  {\bibfnamefont {E.~P. A.~M.}\ \bibnamefont {Bakkers}}, \ and\ \bibinfo
  {author} {\bibfnamefont {L.~P.}\ \bibnamefont {Kouwenhoven}},\ }\href
  {\doibase 10.1126/science.1222360} {\bibfield  {journal} {\bibinfo  {journal}
  {Science}\ }\textbf {\bibinfo {volume} {336}},\ \bibinfo {pages} {1003}
  (\bibinfo {year} {2012})}\BibitemShut {NoStop}%
\bibitem [{\citenamefont {Barkeshli}\ and\ \citenamefont
  {Wen}(2011)}]{Barkeshli11:Genon}%
  \BibitemOpen
  \bibfield  {author} {\bibinfo {author} {\bibfnamefont {M.}~\bibnamefont
  {Barkeshli}}\ and\ \bibinfo {author} {\bibfnamefont {X.-G.}\ \bibnamefont
  {Wen}},\ }\href {\doibase 10.1103/PhysRevB.84.115121} {\bibfield  {journal}
  {\bibinfo  {journal} {Phys. Rev. B}\ }\textbf {\bibinfo {volume} {84}},\
  \bibinfo {pages} {115121} (\bibinfo {year} {2011})}\BibitemShut {NoStop}%
\bibitem [{\citenamefont {Barkeshli}\ and\ \citenamefont
  {Qi}(2012)}]{Barkeshli12:Nematic}%
  \BibitemOpen
  \bibfield  {author} {\bibinfo {author} {\bibfnamefont {M.}~\bibnamefont
  {Barkeshli}}\ and\ \bibinfo {author} {\bibfnamefont {X.-L.}\ \bibnamefont
  {Qi}},\ }\href {\doibase 10.1103/PhysRevX.2.031013} {\bibfield  {journal}
  {\bibinfo  {journal} {Physical Review X}\ }\textbf {\bibinfo {volume} {2}},\
  \bibinfo {pages} {031013} (\bibinfo {year} {2012})}\BibitemShut {NoStop}%
\bibitem [{\citenamefont {Barkeshli}\ \emph
  {et~al.}(2013{\natexlab{a}})\citenamefont {Barkeshli}, \citenamefont {Jian},\
  and\ \citenamefont {Qi}}]{Barkeshli13:Twist}%
  \BibitemOpen
  \bibfield  {author} {\bibinfo {author} {\bibfnamefont {M.}~\bibnamefont
  {Barkeshli}}, \bibinfo {author} {\bibfnamefont {C.-M.}\ \bibnamefont {Jian}},
  \ and\ \bibinfo {author} {\bibfnamefont {X.-L.}\ \bibnamefont {Qi}},\ }\href
  {\doibase 10.1103/PhysRevB.87.045130} {\bibfield  {journal} {\bibinfo
  {journal} {Phys. Rev. B}\ }\textbf {\bibinfo {volume} {87}},\ \bibinfo
  {pages} {045130} (\bibinfo {year} {2013}{\natexlab{a}})}\BibitemShut
  {NoStop}%
\bibitem [{\citenamefont {Clarke}\ \emph {et~al.}(2013)\citenamefont {Clarke},
  \citenamefont {Alicea},\ and\ \citenamefont {Shtengel}}]{Clarke13:FTSC}%
  \BibitemOpen
  \bibfield  {author} {\bibinfo {author} {\bibfnamefont {D.~J.}\ \bibnamefont
  {Clarke}}, \bibinfo {author} {\bibfnamefont {J.}~\bibnamefont {Alicea}}, \
  and\ \bibinfo {author} {\bibfnamefont {K.}~\bibnamefont {Shtengel}},\ }\href
  {\doibase 10.1038/ncomms2340} {\bibfield  {journal} {\bibinfo  {journal}
  {Nature Communications}\ }\textbf {\bibinfo {volume} {4}},\ \bibinfo {pages}
  {1348} (\bibinfo {year} {2013})}\BibitemShut {NoStop}%
\bibitem [{\citenamefont {Lindner}\ \emph {et~al.}(2012)\citenamefont
  {Lindner}, \citenamefont {Berg}, \citenamefont {Refael},\ and\ \citenamefont
  {Stern}}]{Lindner12:FTSC}%
  \BibitemOpen
  \bibfield  {author} {\bibinfo {author} {\bibfnamefont {N.~H.}\ \bibnamefont
  {Lindner}}, \bibinfo {author} {\bibfnamefont {E.}~\bibnamefont {Berg}},
  \bibinfo {author} {\bibfnamefont {G.}~\bibnamefont {Refael}}, \ and\ \bibinfo
  {author} {\bibfnamefont {A.}~\bibnamefont {Stern}},\ }\href {\doibase
  10.1103/PhysRevX.2.041002} {\bibfield  {journal} {\bibinfo  {journal} {Phys.
  Rev. X}\ }\textbf {\bibinfo {volume} {2}},\ \bibinfo {pages} {041002}
  (\bibinfo {year} {2012})}\BibitemShut {NoStop}%
\bibitem [{\citenamefont {Cheng}(2012)}]{Cheng12:FTSC}%
  \BibitemOpen
  \bibfield  {author} {\bibinfo {author} {\bibfnamefont {M.}~\bibnamefont
  {Cheng}},\ }\href {\doibase 10.1103/PhysRevB.86.195126} {\bibfield  {journal}
  {\bibinfo  {journal} {Phys. Rev. B}\ }\textbf {\bibinfo {volume} {86}},\
  \bibinfo {pages} {195126} (\bibinfo {year} {2012})}\BibitemShut {NoStop}%
\bibitem [{\citenamefont {Vaezi}(2013)}]{Vaezi13:FTSC}%
  \BibitemOpen
  \bibfield  {author} {\bibinfo {author} {\bibfnamefont {A.}~\bibnamefont
  {Vaezi}},\ }\href {\doibase 10.1103/PhysRevB.87.035132} {\bibfield  {journal}
  {\bibinfo  {journal} {Phys. Rev. B}\ }\textbf {\bibinfo {volume} {87}},\
  \bibinfo {pages} {035132} (\bibinfo {year} {2013})}\BibitemShut {NoStop}%
\bibitem [{\citenamefont {Barkeshli}\ \emph
  {et~al.}(2013{\natexlab{b}})\citenamefont {Barkeshli}, \citenamefont {Jian},\
  and\ \citenamefont {Qi}}]{Barkeshli13:Defects}%
  \BibitemOpen
  \bibfield  {author} {\bibinfo {author} {\bibfnamefont {M.}~\bibnamefont
  {Barkeshli}}, \bibinfo {author} {\bibfnamefont {C.-M.}\ \bibnamefont {Jian}},
  \ and\ \bibinfo {author} {\bibfnamefont {X.-L.}\ \bibnamefont {Qi}},\ }\href
  {\doibase 10.1103/PhysRevB.88.235103} {\bibfield  {journal} {\bibinfo
  {journal} {Phys. Rev. B}\ }\textbf {\bibinfo {volume} {88}},\ \bibinfo
  {pages} {235103} (\bibinfo {year} {2013}{\natexlab{b}})}\BibitemShut
  {NoStop}%
\bibitem [{\citenamefont {Mong}\ \emph {et~al.}(2014)\citenamefont {Mong},
  \citenamefont {Clarke}, \citenamefont {Alicea}, \citenamefont {Lindner},
  \citenamefont {Fendley}, \citenamefont {Nayak}, \citenamefont {Oreg},
  \citenamefont {Stern}, \citenamefont {Berg}, \citenamefont {Shtengel},\ and\
  \citenamefont {Fisher}}]{Mong14:RR}%
  \BibitemOpen
  \bibfield  {author} {\bibinfo {author} {\bibfnamefont {R.~S.~K.}\
  \bibnamefont {Mong}}, \bibinfo {author} {\bibfnamefont {D.~J.}\ \bibnamefont
  {Clarke}}, \bibinfo {author} {\bibfnamefont {J.}~\bibnamefont {Alicea}},
  \bibinfo {author} {\bibfnamefont {N.~H.}\ \bibnamefont {Lindner}}, \bibinfo
  {author} {\bibfnamefont {P.}~\bibnamefont {Fendley}}, \bibinfo {author}
  {\bibfnamefont {C.}~\bibnamefont {Nayak}}, \bibinfo {author} {\bibfnamefont
  {Y.}~\bibnamefont {Oreg}}, \bibinfo {author} {\bibfnamefont {A.}~\bibnamefont
  {Stern}}, \bibinfo {author} {\bibfnamefont {E.}~\bibnamefont {Berg}},
  \bibinfo {author} {\bibfnamefont {K.}~\bibnamefont {Shtengel}}, \ and\
  \bibinfo {author} {\bibfnamefont {M.~P.~A.}\ \bibnamefont {Fisher}},\ }\href
  {\doibase 10.1103/PhysRevX.4.011036} {\bibfield  {journal} {\bibinfo
  {journal} {Phys. Rev. X}\ }\textbf {\bibinfo {volume} {4}},\ \bibinfo {pages}
  {011036} (\bibinfo {year} {2014})}\BibitemShut {NoStop}%
\bibitem [{\citenamefont {Das~Sarma}\ \emph {et~al.}(2005)\citenamefont
  {Das~Sarma}, \citenamefont {Freedman},\ and\ \citenamefont
  {Nayak}}]{DasSarma05:TQC-MR}%
  \BibitemOpen
  \bibfield  {author} {\bibinfo {author} {\bibfnamefont {S.}~\bibnamefont
  {Das~Sarma}}, \bibinfo {author} {\bibfnamefont {M.}~\bibnamefont {Freedman}},
  \ and\ \bibinfo {author} {\bibfnamefont {C.}~\bibnamefont {Nayak}},\ }\href
  {\doibase 10.1103/PhysRevLett.94.166802} {\bibfield  {journal} {\bibinfo
  {journal} {Phys. Rev. Lett.}\ }\textbf {\bibinfo {volume} {94}},\ \bibinfo
  {pages} {166802} (\bibinfo {year} {2005})}\BibitemShut {NoStop}%
\bibitem [{\citenamefont {Stern}\ and\ \citenamefont
  {Halperin}(2006)}]{Stern06:TQC-MR}%
  \BibitemOpen
  \bibfield  {author} {\bibinfo {author} {\bibfnamefont {A.}~\bibnamefont
  {Stern}}\ and\ \bibinfo {author} {\bibfnamefont {B.~I.}\ \bibnamefont
  {Halperin}},\ }\href {\doibase 10.1103/PhysRevLett.96.016802} {\bibfield
  {journal} {\bibinfo  {journal} {Phys. Rev. Lett.}\ }\textbf {\bibinfo
  {volume} {96}},\ \bibinfo {pages} {016802} (\bibinfo {year}
  {2006})}\BibitemShut {NoStop}%
\bibitem [{\citenamefont {Bonderson}\ \emph {et~al.}(2006)\citenamefont
  {Bonderson}, \citenamefont {Shtengel},\ and\ \citenamefont
  {Slingerland}}]{Bonderson06:TQC-RR}%
  \BibitemOpen
  \bibfield  {author} {\bibinfo {author} {\bibfnamefont {P.}~\bibnamefont
  {Bonderson}}, \bibinfo {author} {\bibfnamefont {K.}~\bibnamefont {Shtengel}},
  \ and\ \bibinfo {author} {\bibfnamefont {J.~K.}\ \bibnamefont
  {Slingerland}},\ }\href {\doibase 10.1103/PhysRevLett.97.016401} {\bibfield
  {journal} {\bibinfo  {journal} {Phys. Rev. Lett.}\ }\textbf {\bibinfo
  {volume} {97}},\ \bibinfo {pages} {016401} (\bibinfo {year}
  {2006})}\BibitemShut {NoStop}%
\bibitem [{\citenamefont {Tsui}\ \emph {et~al.}(1982)\citenamefont {Tsui},
  \citenamefont {Stormer},\ and\ \citenamefont {Gossard}}]{Tsui82:FQH}%
  \BibitemOpen
  \bibfield  {author} {\bibinfo {author} {\bibfnamefont {D.~C.}\ \bibnamefont
  {Tsui}}, \bibinfo {author} {\bibfnamefont {H.~L.}\ \bibnamefont {Stormer}}, \
  and\ \bibinfo {author} {\bibfnamefont {A.~C.}\ \bibnamefont {Gossard}},\
  }\href {\doibase 10.1103/PhysRevLett.48.1559} {\bibfield  {journal} {\bibinfo
   {journal} {Phys. Rev. Lett.}\ }\textbf {\bibinfo {volume} {48}},\ \bibinfo
  {pages} {1559} (\bibinfo {year} {1982})}\BibitemShut {NoStop}%
\bibitem [{\citenamefont {Willett}\ \emph {et~al.}(1987)\citenamefont
  {Willett}, \citenamefont {Eisenstein}, \citenamefont {St{\"{o}}rmer},
  \citenamefont {Tsui}, \citenamefont {Gossard},\ and\ \citenamefont
  {English}}]{Willett87:MR}%
  \BibitemOpen
  \bibfield  {author} {\bibinfo {author} {\bibfnamefont {R.}~\bibnamefont
  {Willett}}, \bibinfo {author} {\bibfnamefont {J.~P.}\ \bibnamefont
  {Eisenstein}}, \bibinfo {author} {\bibfnamefont {H.~L.}\ \bibnamefont
  {St{\"{o}}rmer}}, \bibinfo {author} {\bibfnamefont {D.~C.}\ \bibnamefont
  {Tsui}}, \bibinfo {author} {\bibfnamefont {A.~C.}\ \bibnamefont {Gossard}}, \
  and\ \bibinfo {author} {\bibfnamefont {J.~H.}\ \bibnamefont {English}},\
  }\href {\doibase 10.1103/PhysRevLett.59.1776} {\bibfield  {journal} {\bibinfo
   {journal} {Phys. Rev. Lett.}\ }\textbf {\bibinfo {volume} {59}},\ \bibinfo
  {pages} {1776} (\bibinfo {year} {1987})}\BibitemShut {NoStop}%
\bibitem [{\citenamefont {Pan}\ \emph {et~al.}(1999)\citenamefont {Pan},
  \citenamefont {Xia}, \citenamefont {Shvarts}, \citenamefont {Adams},
  \citenamefont {Stormer}, \citenamefont {Tsui}, \citenamefont {Pfeiffer},
  \citenamefont {Baldwin},\ and\ \citenamefont {West}}]{Pan99:MR}%
  \BibitemOpen
  \bibfield  {author} {\bibinfo {author} {\bibfnamefont {W.}~\bibnamefont
  {Pan}}, \bibinfo {author} {\bibfnamefont {J.~S.}\ \bibnamefont {Xia}},
  \bibinfo {author} {\bibfnamefont {V.}~\bibnamefont {Shvarts}}, \bibinfo
  {author} {\bibfnamefont {D.~E.}\ \bibnamefont {Adams}}, \bibinfo {author}
  {\bibfnamefont {H.~L.}\ \bibnamefont {Stormer}}, \bibinfo {author}
  {\bibfnamefont {D.~C.}\ \bibnamefont {Tsui}}, \bibinfo {author}
  {\bibfnamefont {L.~N.}\ \bibnamefont {Pfeiffer}}, \bibinfo {author}
  {\bibfnamefont {K.~W.}\ \bibnamefont {Baldwin}}, \ and\ \bibinfo {author}
  {\bibfnamefont {K.~W.}\ \bibnamefont {West}},\ }\href {\doibase
  10.1103/PhysRevLett.83.3530} {\bibfield  {journal} {\bibinfo  {journal}
  {Phys. Rev. Lett.}\ }\textbf {\bibinfo {volume} {83}},\ \bibinfo {pages}
  {3530} (\bibinfo {year} {1999})}\BibitemShut {NoStop}%
\bibitem [{\citenamefont {Xia}\ \emph {et~al.}(2004)\citenamefont {Xia},
  \citenamefont {Pan}, \citenamefont {Vicente}, \citenamefont {Adams},
  \citenamefont {Sullivan}, \citenamefont {Stormer}, \citenamefont {Tsui},
  \citenamefont {Pfeiffer}, \citenamefont {Baldwin},\ and\ \citenamefont
  {West}}]{Xia04:RR}%
  \BibitemOpen
  \bibfield  {author} {\bibinfo {author} {\bibfnamefont {J.~S.}\ \bibnamefont
  {Xia}}, \bibinfo {author} {\bibfnamefont {W.}~\bibnamefont {Pan}}, \bibinfo
  {author} {\bibfnamefont {C.~L.}\ \bibnamefont {Vicente}}, \bibinfo {author}
  {\bibfnamefont {E.~D.}\ \bibnamefont {Adams}}, \bibinfo {author}
  {\bibfnamefont {N.~S.}\ \bibnamefont {Sullivan}}, \bibinfo {author}
  {\bibfnamefont {H.~L.}\ \bibnamefont {Stormer}}, \bibinfo {author}
  {\bibfnamefont {D.~C.}\ \bibnamefont {Tsui}}, \bibinfo {author}
  {\bibfnamefont {L.~N.}\ \bibnamefont {Pfeiffer}}, \bibinfo {author}
  {\bibfnamefont {K.~W.}\ \bibnamefont {Baldwin}}, \ and\ \bibinfo {author}
  {\bibfnamefont {K.~W.}\ \bibnamefont {West}},\ }\href {\doibase
  10.1103/PhysRevLett.93.176809} {\bibfield  {journal} {\bibinfo  {journal}
  {Phys. Rev. Lett.}\ }\textbf {\bibinfo {volume} {93}},\ \bibinfo {pages}
  {176809} (\bibinfo {year} {2004})}\BibitemShut {NoStop}%
\bibitem [{\citenamefont {Moore}\ and\ \citenamefont
  {Read}(1991)}]{Moore91:MR}%
  \BibitemOpen
  \bibfield  {author} {\bibinfo {author} {\bibfnamefont {G.}~\bibnamefont
  {Moore}}\ and\ \bibinfo {author} {\bibfnamefont {N.}~\bibnamefont {Read}},\
  }\href {\doibase 10.1016/0550-3213(91)90407-O} {\bibfield  {journal}
  {\bibinfo  {journal} {Nuclear Physics B}\ }\textbf {\bibinfo {volume}
  {360}},\ \bibinfo {pages} {362} (\bibinfo {year} {1991})}\BibitemShut
  {NoStop}%
\bibitem [{\citenamefont {Read}\ and\ \citenamefont
  {Rezayi}(1999)}]{Read99:RR}%
  \BibitemOpen
  \bibfield  {author} {\bibinfo {author} {\bibfnamefont {N.}~\bibnamefont
  {Read}}\ and\ \bibinfo {author} {\bibfnamefont {E.}~\bibnamefont {Rezayi}},\
  }\href {\doibase 10.1103/PhysRevB.59.8084} {\bibfield  {journal} {\bibinfo
  {journal} {Physical Review B}\ }\textbf {\bibinfo {volume} {59}},\ \bibinfo
  {pages} {8084} (\bibinfo {year} {1999})}\BibitemShut {NoStop}%
\bibitem [{\citenamefont {Fubini}(1991)}]{Fubini91:FQH-CFT}%
  \BibitemOpen
  \bibfield  {author} {\bibinfo {author} {\bibfnamefont {S.}~\bibnamefont
  {Fubini}},\ }\href {\doibase 10.1142/S0217732391000336} {\bibfield  {journal}
  {\bibinfo  {journal} {Modern Physics Letters A}\ }\textbf {\bibinfo {volume}
  {06}},\ \bibinfo {pages} {347} (\bibinfo {year} {1991})}\BibitemShut
  {NoStop}%
\bibitem [{\citenamefont {Belavin}\ \emph {et~al.}(1984)\citenamefont
  {Belavin}, \citenamefont {Polyakov},\ and\ \citenamefont
  {Zamolodchikov}}]{Belavin84:BPZ}%
  \BibitemOpen
  \bibfield  {author} {\bibinfo {author} {\bibfnamefont {A.~A.}\ \bibnamefont
  {Belavin}}, \bibinfo {author} {\bibfnamefont {A.~M.}\ \bibnamefont
  {Polyakov}}, \ and\ \bibinfo {author} {\bibfnamefont {A.~B.}\ \bibnamefont
  {Zamolodchikov}},\ }\href {\doibase
  http://dx.doi.org/10.1016/0550-3213(84)90052-X} {\bibfield  {journal}
  {\bibinfo  {journal} {Nuclear Physics B}\ }\textbf {\bibinfo {volume}
  {241}},\ \bibinfo {pages} {333} (\bibinfo {year} {1984})}\BibitemShut
  {NoStop}%
\bibitem [{\citenamefont {Moore}\ and\ \citenamefont
  {Seiberg}(1989)}]{Moore89:CFT}%
  \BibitemOpen
  \bibfield  {author} {\bibinfo {author} {\bibfnamefont {G.}~\bibnamefont
  {Moore}}\ and\ \bibinfo {author} {\bibfnamefont {N.}~\bibnamefont
  {Seiberg}},\ }\href {\doibase 10.1007/BF01238857} {\bibfield  {journal}
  {\bibinfo  {journal} {Communications in Mathematical Physics}\ }\textbf
  {\bibinfo {volume} {123}},\ \bibinfo {pages} {177} (\bibinfo {year}
  {1989})}\BibitemShut {NoStop}%
\bibitem [{\citenamefont {{Di Francesco}}\ \emph {et~al.}(1999)\citenamefont
  {{Di Francesco}}, \citenamefont {Mathieu},\ and\ \citenamefont
  {S{\'e}n{\'e}chal}}]{DiFrancesco99:Yellow}%
  \BibitemOpen
  \bibfield  {author} {\bibinfo {author} {\bibfnamefont {P.}~\bibnamefont {{Di
  Francesco}}}, \bibinfo {author} {\bibfnamefont {P.}~\bibnamefont {Mathieu}},
  \ and\ \bibinfo {author} {\bibfnamefont {D.}~\bibnamefont
  {S{\'e}n{\'e}chal}},\ }\href@noop {} {\emph {\bibinfo {title} {Conformal
  Field Theory}}}\ (\bibinfo  {publisher} {Springer},\ \bibinfo {year}
  {1999})\BibitemShut {NoStop}%
\bibitem [{\citenamefont {Laughlin}(1983)}]{Laughlin83:Nobel}%
  \BibitemOpen
  \bibfield  {author} {\bibinfo {author} {\bibfnamefont {R.~B.}\ \bibnamefont
  {Laughlin}},\ }\href {\doibase 10.1103/PhysRevLett.50.1395} {\bibfield
  {journal} {\bibinfo  {journal} {Physical Review Letters}\ }\textbf {\bibinfo
  {volume} {50}},\ \bibinfo {pages} {1395} (\bibinfo {year}
  {1983})}\BibitemShut {NoStop}%
\bibitem [{\citenamefont {Arovas}\ \emph {et~al.}(1984)\citenamefont {Arovas},
  \citenamefont {Schrieffer},\ and\ \citenamefont
  {Wilczek}}]{Arovas84:Statistics}%
  \BibitemOpen
  \bibfield  {author} {\bibinfo {author} {\bibfnamefont {D.}~\bibnamefont
  {Arovas}}, \bibinfo {author} {\bibfnamefont {J.~R.}\ \bibnamefont
  {Schrieffer}}, \ and\ \bibinfo {author} {\bibfnamefont {F.}~\bibnamefont
  {Wilczek}},\ }\href {\doibase 10.1103/PhysRevLett.53.722} {\bibfield
  {journal} {\bibinfo  {journal} {Phys. Rev. Lett.}\ }\textbf {\bibinfo
  {volume} {53}},\ \bibinfo {pages} {722} (\bibinfo {year} {1984})}\BibitemShut
  {NoStop}%
\bibitem [{\citenamefont {Gurarie}\ and\ \citenamefont
  {Nayak}(1997)}]{Gurarie97:Plasma}%
  \BibitemOpen
  \bibfield  {author} {\bibinfo {author} {\bibfnamefont {V.}~\bibnamefont
  {Gurarie}}\ and\ \bibinfo {author} {\bibfnamefont {C.}~\bibnamefont
  {Nayak}},\ }\href {\doibase http://dx.doi.org/10.1016/S0550-3213(97)00612-3}
  {\bibfield  {journal} {\bibinfo  {journal} {Nuclear Physics B}\ }\textbf
  {\bibinfo {volume} {506}},\ \bibinfo {pages} {685} (\bibinfo {year}
  {1997})}\BibitemShut {NoStop}%
\bibitem [{\citenamefont {Read}(2009)}]{Read09:Adiabatic}%
  \BibitemOpen
  \bibfield  {author} {\bibinfo {author} {\bibfnamefont {N.}~\bibnamefont
  {Read}},\ }\href {\doibase 10.1103/PhysRevB.79.045308} {\bibfield  {journal}
  {\bibinfo  {journal} {Phys. Rev. B}\ }\textbf {\bibinfo {volume} {79}},\
  \bibinfo {pages} {045308} (\bibinfo {year} {2009})}\BibitemShut {NoStop}%
\bibitem [{\citenamefont {Bonderson}\ \emph {et~al.}(2011)\citenamefont
  {Bonderson}, \citenamefont {Gurarie},\ and\ \citenamefont
  {Nayak}}]{Bonderson11:Plasma}%
  \BibitemOpen
  \bibfield  {author} {\bibinfo {author} {\bibfnamefont {P.}~\bibnamefont
  {Bonderson}}, \bibinfo {author} {\bibfnamefont {V.}~\bibnamefont {Gurarie}},
  \ and\ \bibinfo {author} {\bibfnamefont {C.}~\bibnamefont {Nayak}},\ }\href
  {\doibase 10.1103/PhysRevB.83.075303} {\bibfield  {journal} {\bibinfo
  {journal} {Physical Review B}\ }\textbf {\bibinfo {volume} {83}},\ \bibinfo
  {pages} {075303} (\bibinfo {year} {2011})}\BibitemShut {NoStop}%
\bibitem [{\citenamefont {Tserkovnyak}\ and\ \citenamefont
  {Simon}(2003)}]{Tserkovnyak03:MR}%
  \BibitemOpen
  \bibfield  {author} {\bibinfo {author} {\bibfnamefont {Y.}~\bibnamefont
  {Tserkovnyak}}\ and\ \bibinfo {author} {\bibfnamefont {S.~H.}\ \bibnamefont
  {Simon}},\ }\href {\doibase 10.1103/PhysRevLett.90.016802} {\bibfield
  {journal} {\bibinfo  {journal} {Phys. Rev. Lett.}\ }\textbf {\bibinfo
  {volume} {90}},\ \bibinfo {pages} {016802} (\bibinfo {year}
  {2003})}\BibitemShut {NoStop}%
\bibitem [{\citenamefont {Prodan}\ and\ \citenamefont
  {Haldane}(2009)}]{Prodan09:Braiding}%
  \BibitemOpen
  \bibfield  {author} {\bibinfo {author} {\bibfnamefont {E.}~\bibnamefont
  {Prodan}}\ and\ \bibinfo {author} {\bibfnamefont {F.~D.~M.}\ \bibnamefont
  {Haldane}},\ }\href {\doibase 10.1103/PhysRevB.80.115121} {\bibfield
  {journal} {\bibinfo  {journal} {Physical Review B}\ }\textbf {\bibinfo
  {volume} {80}},\ \bibinfo {pages} {115121} (\bibinfo {year}
  {2009})}\BibitemShut {NoStop}%
\bibitem [{\citenamefont {Baraban}\ \emph {et~al.}(2009)\citenamefont
  {Baraban}, \citenamefont {Zikos}, \citenamefont {Bonesteel},\ and\
  \citenamefont {Simon}}]{Baraban09:MR}%
  \BibitemOpen
  \bibfield  {author} {\bibinfo {author} {\bibfnamefont {M.}~\bibnamefont
  {Baraban}}, \bibinfo {author} {\bibfnamefont {G.}~\bibnamefont {Zikos}},
  \bibinfo {author} {\bibfnamefont {N.}~\bibnamefont {Bonesteel}}, \ and\
  \bibinfo {author} {\bibfnamefont {S.~H.}\ \bibnamefont {Simon}},\ }\href
  {\doibase 10.1103/PhysRevLett.103.076801} {\bibfield  {journal} {\bibinfo
  {journal} {Phys. Rev. Lett.}\ }\textbf {\bibinfo {volume} {103}},\ \bibinfo
  {pages} {076801} (\bibinfo {year} {2009})}\BibitemShut {NoStop}%
\bibitem [{\citenamefont {Simon}\ \emph {et~al.}(2007)\citenamefont {Simon},
  \citenamefont {Rezayi}, \citenamefont {Cooper},\ and\ \citenamefont
  {Berdnikov}}]{Simon07:Gaffnian}%
  \BibitemOpen
  \bibfield  {author} {\bibinfo {author} {\bibfnamefont {S.~H.}\ \bibnamefont
  {Simon}}, \bibinfo {author} {\bibfnamefont {E.~H.}\ \bibnamefont {Rezayi}},
  \bibinfo {author} {\bibfnamefont {N.~R.}\ \bibnamefont {Cooper}}, \ and\
  \bibinfo {author} {\bibfnamefont {I.}~\bibnamefont {Berdnikov}},\ }\href
  {\doibase 10.1103/PhysRevB.75.075317} {\bibfield  {journal} {\bibinfo
  {journal} {Physical Review B}\ }\textbf {\bibinfo {volume} {75}},\ \bibinfo
  {pages} {075317} (\bibinfo {year} {2007})}\BibitemShut {NoStop}%
\bibitem [{\citenamefont {Estienne}\ \emph {et~al.}(2014)\citenamefont
  {Estienne}, \citenamefont {Regnault},\ and\ \citenamefont
  {Bernevig}}]{Estienne14:Gaffnian}%
  \BibitemOpen
  \bibfield  {author} {\bibinfo {author} {\bibfnamefont {B.}~\bibnamefont
  {Estienne}}, \bibinfo {author} {\bibfnamefont {N.}~\bibnamefont {Regnault}},
  \ and\ \bibinfo {author} {\bibfnamefont {B.~A.}\ \bibnamefont {Bernevig}},\
  }\href@noop {} {\bibfield  {journal} {\bibinfo  {journal} {ArXiv e-prints}\ }
  (\bibinfo {year} {2014})},\ \Eprint {http://arxiv.org/abs/1406.6262}
  {arXiv:1406.6262 [cond-mat.str-el]} \BibitemShut {NoStop}%
\bibitem [{\citenamefont {Dubail}\ \emph {et~al.}(2012)\citenamefont {Dubail},
  \citenamefont {Read},\ and\ \citenamefont {Rezayi}}]{Dubail12:MPS}%
  \BibitemOpen
  \bibfield  {author} {\bibinfo {author} {\bibfnamefont {J.}~\bibnamefont
  {Dubail}}, \bibinfo {author} {\bibfnamefont {N.}~\bibnamefont {Read}}, \ and\
  \bibinfo {author} {\bibfnamefont {E.~H.}\ \bibnamefont {Rezayi}},\ }\href
  {\doibase 10.1103/PhysRevB.86.245310} {\bibfield  {journal} {\bibinfo
  {journal} {Physical Review B}\ }\textbf {\bibinfo {volume} {86}},\ \bibinfo
  {pages} {245310} (\bibinfo {year} {2012})}\BibitemShut {NoStop}%
\bibitem [{\citenamefont {Zaletel}\ and\ \citenamefont
  {Mong}(2012)}]{Zaletel12:MPS}%
  \BibitemOpen
  \bibfield  {author} {\bibinfo {author} {\bibfnamefont {M.~P.}\ \bibnamefont
  {Zaletel}}\ and\ \bibinfo {author} {\bibfnamefont {R.~S.~K.}\ \bibnamefont
  {Mong}},\ }\href {\doibase 10.1103/PhysRevB.86.245305} {\bibfield  {journal}
  {\bibinfo  {journal} {Physical Review B}\ }\textbf {\bibinfo {volume} {86}},\
  \bibinfo {pages} {245305} (\bibinfo {year} {2012})}\BibitemShut {NoStop}%
\bibitem [{\citenamefont {Estienne}\ \emph
  {et~al.}(2013{\natexlab{a}})\citenamefont {Estienne}, \citenamefont {Papic},
  \citenamefont {Regnault},\ and\ \citenamefont {Bernevig}}]{Estienne13:MPS}%
  \BibitemOpen
  \bibfield  {author} {\bibinfo {author} {\bibfnamefont {B.}~\bibnamefont
  {Estienne}}, \bibinfo {author} {\bibfnamefont {Z.}~\bibnamefont {Papic}},
  \bibinfo {author} {\bibfnamefont {N.}~\bibnamefont {Regnault}}, \ and\
  \bibinfo {author} {\bibfnamefont {B.~A.}\ \bibnamefont {Bernevig}},\ }\href
  {\doibase 10.1103/PhysRevB.87.161112} {\bibfield  {journal} {\bibinfo
  {journal} {Physical Review B}\ }\textbf {\bibinfo {volume} {87}},\ \bibinfo
  {pages} {161112} (\bibinfo {year} {2013}{\natexlab{a}})}\BibitemShut
  {NoStop}%
\bibitem [{\citenamefont {Estienne}\ \emph
  {et~al.}(2013{\natexlab{b}})\citenamefont {Estienne}, \citenamefont
  {Regnault},\ and\ \citenamefont {Bernevig}}]{Estienne13:MPSLong}%
  \BibitemOpen
  \bibfield  {author} {\bibinfo {author} {\bibfnamefont {B.}~\bibnamefont
  {Estienne}}, \bibinfo {author} {\bibfnamefont {N.}~\bibnamefont {Regnault}},
  \ and\ \bibinfo {author} {\bibfnamefont {B.~A.}\ \bibnamefont {Bernevig}},\
  }\href@noop {} {\bibfield  {journal} {\bibinfo  {journal} {ArXiv e-prints}\ }
  (\bibinfo {year} {2013}{\natexlab{b}})},\ \Eprint
  {http://arxiv.org/abs/1311.2936} {arXiv:1311.2936 [cond-mat.str-el]}
  \BibitemShut {NoStop}%
\bibitem [{\citenamefont {Fannes}\ \emph {et~al.}(1992)\citenamefont {Fannes},
  \citenamefont {Nachtergaele},\ and\ \citenamefont {Werner}}]{Fannes92:MPS}%
  \BibitemOpen
  \bibfield  {author} {\bibinfo {author} {\bibfnamefont {M.}~\bibnamefont
  {Fannes}}, \bibinfo {author} {\bibfnamefont {B.}~\bibnamefont
  {Nachtergaele}}, \ and\ \bibinfo {author} {\bibfnamefont {R.~F.}\
  \bibnamefont {Werner}},\ }\href {\doibase 10.1007/BF02099178} {\bibfield
  {journal} {\bibinfo  {journal} {Communications in Mathematical Physics}\
  }\textbf {\bibinfo {volume} {144}},\ \bibinfo {pages} {443} (\bibinfo {year}
  {1992})}\BibitemShut {NoStop}%
\bibitem [{\citenamefont {Schollw{\"{o}}ck}(2011)}]{Schollwock11:DMRG-MPS}%
  \BibitemOpen
  \bibfield  {author} {\bibinfo {author} {\bibfnamefont {U.}~\bibnamefont
  {Schollw{\"{o}}ck}},\ }\href {\doibase 10.1016/j.aop.2010.09.012} {\bibfield
  {journal} {\bibinfo  {journal} {Annals of Physics}\ }\textbf {\bibinfo
  {volume} {326}},\ \bibinfo {pages} {96} (\bibinfo {year} {2011})}\BibitemShut
  {NoStop}%
\bibitem [{\citenamefont {Wu}\ \emph {et~al.}(2014)\citenamefont {Wu},
  \citenamefont {Estienne}, \citenamefont {Regnault},\ and\ \citenamefont
  {Bernevig}}]{Wu14:MPS}%
  \BibitemOpen
  \bibfield  {author} {\bibinfo {author} {\bibfnamefont {Y.-L.}\ \bibnamefont
  {Wu}}, \bibinfo {author} {\bibfnamefont {B.}~\bibnamefont {Estienne}},
  \bibinfo {author} {\bibfnamefont {N.}~\bibnamefont {Regnault}}, \ and\
  \bibinfo {author} {\bibfnamefont {B.~A.}\ \bibnamefont {Bernevig}},\
  }\href@noop {} {} (\bibinfo {year} {2014}),\ \bibinfo {note} {in
  preparation}\BibitemShut {NoStop}%
\bibitem [{\citenamefont {Rezayi}\ and\ \citenamefont
  {Haldane}(1994)}]{Rezayi94:Cylinder}%
  \BibitemOpen
  \bibfield  {author} {\bibinfo {author} {\bibfnamefont {E.~H.}\ \bibnamefont
  {Rezayi}}\ and\ \bibinfo {author} {\bibfnamefont {F.~D.~M.}\ \bibnamefont
  {Haldane}},\ }\href {\doibase 10.1103/PhysRevB.50.17199} {\bibfield
  {journal} {\bibinfo  {journal} {Phys. Rev. B}\ }\textbf {\bibinfo {volume}
  {50}},\ \bibinfo {pages} {17199} (\bibinfo {year} {1994})}\BibitemShut
  {NoStop}%
\bibitem [{\citenamefont {Yurov}\ and\ \citenamefont
  {Zamolodchikov}(1990)}]{Yurov90:TCSA}%
  \BibitemOpen
  \bibfield  {author} {\bibinfo {author} {\bibfnamefont {V.~P.}\ \bibnamefont
  {Yurov}}\ and\ \bibinfo {author} {\bibfnamefont {A.~B.}\ \bibnamefont
  {Zamolodchikov}},\ }\href {\doibase 10.1142/S0217751X9000218X} {\bibfield
  {journal} {\bibinfo  {journal} {International Journal of Modern Physics A}\
  }\textbf {\bibinfo {volume} {05}},\ \bibinfo {pages} {3221} (\bibinfo {year}
  {1990})}\BibitemShut {NoStop}%
\bibitem [{\citenamefont {Eisert}\ \emph {et~al.}(2010)\citenamefont {Eisert},
  \citenamefont {Cramer},\ and\ \citenamefont {Plenio}}]{Eisert10:AreaLaw}%
  \BibitemOpen
  \bibfield  {author} {\bibinfo {author} {\bibfnamefont {J.}~\bibnamefont
  {Eisert}}, \bibinfo {author} {\bibfnamefont {M.}~\bibnamefont {Cramer}}, \
  and\ \bibinfo {author} {\bibfnamefont {M.~B.}\ \bibnamefont {Plenio}},\
  }\href {\doibase 10.1103/RevModPhys.82.277} {\bibfield  {journal} {\bibinfo
  {journal} {Rev. Mod. Phys.}\ }\textbf {\bibinfo {volume} {82}},\ \bibinfo
  {pages} {277} (\bibinfo {year} {2010})}\BibitemShut {NoStop}%
\bibitem [{\citenamefont {Nayak}\ and\ \citenamefont
  {Wilczek}(1996)}]{Nayak96:SO2n}%
  \BibitemOpen
  \bibfield  {author} {\bibinfo {author} {\bibfnamefont {C.}~\bibnamefont
  {Nayak}}\ and\ \bibinfo {author} {\bibfnamefont {F.}~\bibnamefont
  {Wilczek}},\ }\href {\doibase http://dx.doi.org/10.1016/0550-3213(96)00430-0}
  {\bibfield  {journal} {\bibinfo  {journal} {Nuclear Physics B}\ }\textbf
  {\bibinfo {volume} {479}},\ \bibinfo {pages} {529} (\bibinfo {year}
  {1996})}\BibitemShut {NoStop}%
\bibitem [{\citenamefont {Kitaev}(2006)}]{Kitaev06:Anyon}%
  \BibitemOpen
  \bibfield  {author} {\bibinfo {author} {\bibfnamefont {A.}~\bibnamefont
  {Kitaev}},\ }\href {\doibase 10.1016/j.aop.2005.10.005} {\bibfield  {journal}
  {\bibinfo  {journal} {Annals of Physics}\ }\textbf {\bibinfo {volume}
  {321}},\ \bibinfo {pages} {2} (\bibinfo {year} {2006})}\BibitemShut {NoStop}%
\bibitem [{Note1()}]{Note1}%
  \BibitemOpen
  \bibinfo {note} {We leave the compactified U(1) boson implicit.}\BibitemShut
  {Stop}%
\bibitem [{\citenamefont {Ardonne}\ and\ \citenamefont
  {Schoutens}(2007)}]{Ardonne07:RR}%
  \BibitemOpen
  \bibfield  {author} {\bibinfo {author} {\bibfnamefont {E.}~\bibnamefont
  {Ardonne}}\ and\ \bibinfo {author} {\bibfnamefont {K.}~\bibnamefont
  {Schoutens}},\ }\href {\doibase http://dx.doi.org/10.1016/j.aop.2006.07.015}
  {\bibfield  {journal} {\bibinfo  {journal} {Annals of Physics}\ }\textbf
  {\bibinfo {volume} {322}},\ \bibinfo {pages} {201} (\bibinfo {year}
  {2007})}\BibitemShut {NoStop}%
\bibitem [{Wu1()}]{Wu14:Supplemental}%
  \BibitemOpen
  \href@noop {} {}\bibinfo {note} {Supplemental Material}\BibitemShut {NoStop}%
\bibitem [{\citenamefont {Johri}\ \emph {et~al.}(2014)\citenamefont {Johri},
  \citenamefont {Papi{\'c}}, \citenamefont {Bhatt},\ and\ \citenamefont
  {Schmitteckert}}]{Johri14:Quasihole}%
  \BibitemOpen
  \bibfield  {author} {\bibinfo {author} {\bibfnamefont {S.}~\bibnamefont
  {Johri}}, \bibinfo {author} {\bibfnamefont {Z.}~\bibnamefont {Papi{\'c}}},
  \bibinfo {author} {\bibfnamefont {R.~N.}\ \bibnamefont {Bhatt}}, \ and\
  \bibinfo {author} {\bibfnamefont {P.}~\bibnamefont {Schmitteckert}},\ }\href
  {\doibase 10.1103/PhysRevB.89.115124} {\bibfield  {journal} {\bibinfo
  {journal} {Phys. Rev. B}\ }\textbf {\bibinfo {volume} {89}},\ \bibinfo
  {pages} {115124} (\bibinfo {year} {2014})}\BibitemShut {NoStop}%
\bibitem [{Note2()}]{Note2}%
  \BibitemOpen
  \bibinfo {note} {This definition of $R$ is very sensitive to the tail of the
  charge excess distribution. It clearly distinguishes quasihole sizes in
  different theories, even though the extent of the quasihole seems comparable
  between different curves in Fig.~\ref {fig:density-radius}(a) by direct
  inspection.}\BibitemShut {Stop}%
\bibitem [{\citenamefont {Tao}\ and\ \citenamefont
  {Thouless}(1983)}]{Tao83:TT}%
  \BibitemOpen
  \bibfield  {author} {\bibinfo {author} {\bibfnamefont {R.}~\bibnamefont
  {Tao}}\ and\ \bibinfo {author} {\bibfnamefont {D.~J.}\ \bibnamefont
  {Thouless}},\ }\href {\doibase 10.1103/PhysRevB.28.1142} {\bibfield
  {journal} {\bibinfo  {journal} {Phys. Rev. B}\ }\textbf {\bibinfo {volume}
  {28}},\ \bibinfo {pages} {1142} (\bibinfo {year} {1983})}\BibitemShut
  {NoStop}%
\bibitem [{\citenamefont {Storni}\ and\ \citenamefont
  {Morf}(2011)}]{Storni11:MR}%
  \BibitemOpen
  \bibfield  {author} {\bibinfo {author} {\bibfnamefont {M.}~\bibnamefont
  {Storni}}\ and\ \bibinfo {author} {\bibfnamefont {R.~H.}\ \bibnamefont
  {Morf}},\ }\href {\doibase 10.1103/PhysRevB.83.195306} {\bibfield  {journal}
  {\bibinfo  {journal} {Phys. Rev. B}\ }\textbf {\bibinfo {volume} {83}},\
  \bibinfo {pages} {195306} (\bibinfo {year} {2011})}\BibitemShut {NoStop}%
\bibitem [{Note3()}]{Note3}%
  \BibitemOpen
  \bibinfo {note} {The RR matrices here are unitarily similar to Ref.~\cite
  {Ardonne07:RR}, due to their fusion tree choice $\begin {minipage}[c]{0.7cm}
  {\protect \includegraphics [width=\linewidth ]{fusion-tree-eddy.pdf}}
  \end {minipage}$ versus our $\begin {minipage}[c]{0.7cm} {\protect
  \includegraphics [width=\linewidth ]{fusion-tree-bare.pdf}} \end
  {minipage}$.}\BibitemShut {Stop}%
\bibitem [{Note4()}]{Note4}%
  \BibitemOpen
  \bibinfo {note} {In the actual calculations, we push the inert quasiholes
  outside of the braiding loop to infinity [see Eq.~\protect \textup {\hbox
  {\mathsurround \z@ \protect \normalfont (\ignorespaces \ref
  {eq:fusion-abc}\unskip \@@italiccorr )}}]. This essentially decouples the two
  conformal blocks for the $\protect \{12\protect \}$ braid, leading to very
  accurate (machine precision) diagonal matrices for $\protect \mathcal
  {B}^{\protect \{12\protect \}}$ at moderate $L_y$.}\BibitemShut {Stop}%
\bibitem [{\citenamefont {Rosenow}\ and\ \citenamefont
  {Simon}(2012)}]{Rosenow12:Interferometer}%
  \BibitemOpen
  \bibfield  {author} {\bibinfo {author} {\bibfnamefont {B.}~\bibnamefont
  {Rosenow}}\ and\ \bibinfo {author} {\bibfnamefont {S.~H.}\ \bibnamefont
  {Simon}},\ }\href {\doibase 10.1103/PhysRevB.85.201302} {\bibfield  {journal}
  {\bibinfo  {journal} {Phys. Rev. B}\ }\textbf {\bibinfo {volume} {85}},\
  \bibinfo {pages} {201302} (\bibinfo {year} {2012})}\BibitemShut {NoStop}%
\bibitem [{\citenamefont {Ardonne}\ \emph {et~al.}(2011)\citenamefont
  {Ardonne}, \citenamefont {Gukelberger}, \citenamefont {Ludwig}, \citenamefont
  {Trebst},\ and\ \citenamefont {Troyer}}]{Ardonne11:YangLee}%
  \BibitemOpen
  \bibfield  {author} {\bibinfo {author} {\bibfnamefont {E.}~\bibnamefont
  {Ardonne}}, \bibinfo {author} {\bibfnamefont {J.}~\bibnamefont
  {Gukelberger}}, \bibinfo {author} {\bibfnamefont {A.~W.~W.}\ \bibnamefont
  {Ludwig}}, \bibinfo {author} {\bibfnamefont {S.}~\bibnamefont {Trebst}}, \
  and\ \bibinfo {author} {\bibfnamefont {M.}~\bibnamefont {Troyer}},\ }\href
  {\doibase 10.1088/1367-2630/13/4/045006} {\bibfield  {journal} {\bibinfo
  {journal} {New Journal of Physics}\ }\textbf {\bibinfo {volume} {13}},\
  \bibinfo {pages} {045006} (\bibinfo {year} {2011})}\BibitemShut {NoStop}%
\bibitem [{\citenamefont {Cheng}\ \emph {et~al.}(2009)\citenamefont {Cheng},
  \citenamefont {Lutchyn}, \citenamefont {Galitski},\ and\ \citenamefont
  {Das~Sarma}}]{Cheng09:Splitting}%
  \BibitemOpen
  \bibfield  {author} {\bibinfo {author} {\bibfnamefont {M.}~\bibnamefont
  {Cheng}}, \bibinfo {author} {\bibfnamefont {R.~M.}\ \bibnamefont {Lutchyn}},
  \bibinfo {author} {\bibfnamefont {V.}~\bibnamefont {Galitski}}, \ and\
  \bibinfo {author} {\bibfnamefont {S.}~\bibnamefont {Das~Sarma}},\ }\href
  {\doibase 10.1103/PhysRevLett.103.107001} {\bibfield  {journal} {\bibinfo
  {journal} {Phys. Rev. Lett.}\ }\textbf {\bibinfo {volume} {103}},\ \bibinfo
  {pages} {107001} (\bibinfo {year} {2009})}\BibitemShut {NoStop}%
\bibitem [{\citenamefont {Bonderson}(2009)}]{Bonderson09:Splitting}%
  \BibitemOpen
  \bibfield  {author} {\bibinfo {author} {\bibfnamefont {P.}~\bibnamefont
  {Bonderson}},\ }\href {\doibase 10.1103/PhysRevLett.103.110403} {\bibfield
  {journal} {\bibinfo  {journal} {Phys. Rev. Lett.}\ }\textbf {\bibinfo
  {volume} {103}},\ \bibinfo {pages} {110403} (\bibinfo {year}
  {2009})}\BibitemShut {NoStop}%
\end{thebibliography}

\begin{thebibliography}{7}%
\makeatletter
\providecommand \@ifxundefined [1]{%
 \@ifx{#1\undefined}
}%
\providecommand \@ifnum [1]{%
 \ifnum #1\expandafter \@firstoftwo
 \else \expandafter \@secondoftwo
 \fi
}%
\providecommand \@ifx [1]{%
 \ifx #1\expandafter \@firstoftwo
 \else \expandafter \@secondoftwo
 \fi
}%
\providecommand \natexlab [1]{#1}%
\providecommand \enquote  [1]{``#1''}%
\providecommand \bibnamefont  [1]{#1}%
\providecommand \bibfnamefont [1]{#1}%
\providecommand \citenamefont [1]{#1}%
\providecommand \href@noop [0]{\@secondoftwo}%
\providecommand \href [0]{\begingroup \@sanitize@url \@href}%
\providecommand \@href[1]{\@@startlink{#1}\@@href}%
\providecommand \@@href[1]{\endgroup#1\@@endlink}%
\providecommand \@sanitize@url [0]{\catcode `\\12\catcode `\$12\catcode
  `\&12\catcode `\#12\catcode `\^12\catcode `\_12\catcode `\%12\relax}%
\providecommand \@@startlink[1]{}%
\providecommand \@@endlink[0]{}%
\providecommand \url  [0]{\begingroup\@sanitize@url \@url }%
\providecommand \@url [1]{\endgroup\@href {#1}{\urlprefix }}%
\providecommand \urlprefix  [0]{URL }%
\providecommand \Eprint [0]{\href }%
\providecommand \doibase [0]{http://dx.doi.org/}%
\providecommand \selectlanguage [0]{\@gobble}%
\providecommand \bibinfo  [0]{\@secondoftwo}%
\providecommand \bibfield  [0]{\@secondoftwo}%
\providecommand \translation [1]{[#1]}%
\providecommand \BibitemOpen [0]{}%
\providecommand \bibitemStop [0]{}%
\providecommand \bibitemNoStop [0]{.\EOS\space}%
\providecommand \EOS [0]{\spacefactor3000\relax}%
\providecommand \BibitemShut  [1]{\csname bibitem#1\endcsname}%
\let\auto@bib@innerbib\@empty
\bibitem [{\citenamefont {Read}\ and\ \citenamefont
  {Rezayi}(1999)}]{Read99:RR:Supplemental}%
  \BibitemOpen
  \bibfield  {author} {\bibinfo {author} {\bibfnamefont {N.}~\bibnamefont
  {Read}}\ and\ \bibinfo {author} {\bibfnamefont {E.}~\bibnamefont {Rezayi}},\
  }\href {\doibase 10.1103/PhysRevB.59.8084} {\bibfield  {journal} {\bibinfo
  {journal} {Physical Review B}\ }\textbf {\bibinfo {volume} {59}},\ \bibinfo
  {pages} {8084} (\bibinfo {year} {1999})}\BibitemShut {NoStop}%
\bibitem [{\citenamefont {{Di Francesco}}\ \emph {et~al.}(1999)\citenamefont
  {{Di Francesco}}, \citenamefont {Mathieu},\ and\ \citenamefont
  {S{\'e}n{\'e}chal}}]{DiFrancesco99:Yellow:Supplemental}%
  \BibitemOpen
  \bibfield  {author} {\bibinfo {author} {\bibfnamefont {P.}~\bibnamefont {{Di
  Francesco}}}, \bibinfo {author} {\bibfnamefont {P.}~\bibnamefont {Mathieu}},
  \ and\ \bibinfo {author} {\bibfnamefont {D.}~\bibnamefont
  {S{\'e}n{\'e}chal}},\ }\href@noop {} {\emph {\bibinfo {title} {Conformal
  Field Theory}}}\ (\bibinfo  {publisher} {Springer},\ \bibinfo {year}
  {1999})\BibitemShut {NoStop}%
\bibitem [{\citenamefont {Simon}\ \emph {et~al.}(2007)\citenamefont {Simon},
  \citenamefont {Rezayi}, \citenamefont {Cooper},\ and\ \citenamefont
  {Berdnikov}}]{Simon07:Gaffnian:Supplemental}%
  \BibitemOpen
  \bibfield  {author} {\bibinfo {author} {\bibfnamefont {S.~H.}\ \bibnamefont
  {Simon}}, \bibinfo {author} {\bibfnamefont {E.~H.}\ \bibnamefont {Rezayi}},
  \bibinfo {author} {\bibfnamefont {N.~R.}\ \bibnamefont {Cooper}}, \ and\
  \bibinfo {author} {\bibfnamefont {I.}~\bibnamefont {Berdnikov}},\ }\href
  {\doibase 10.1103/PhysRevB.75.075317} {\bibfield  {journal} {\bibinfo
  {journal} {Physical Review B}\ }\textbf {\bibinfo {volume} {75}},\ \bibinfo
  {pages} {075317} (\bibinfo {year} {2007})}\BibitemShut {NoStop}%
\bibitem [{\citenamefont {Estienne}\ \emph {et~al.}(2013)\citenamefont
  {Estienne}, \citenamefont {Regnault},\ and\ \citenamefont
  {Bernevig}}]{Estienne13:MPSLong:Supplemental}%
  \BibitemOpen
  \bibfield  {author} {\bibinfo {author} {\bibfnamefont {B.}~\bibnamefont
  {Estienne}}, \bibinfo {author} {\bibfnamefont {N.}~\bibnamefont {Regnault}},
  \ and\ \bibinfo {author} {\bibfnamefont {B.~A.}\ \bibnamefont {Bernevig}},\
  }\href@noop {} {\bibfield  {journal} {\bibinfo  {journal} {ArXiv e-prints}\ }
  (\bibinfo {year} {2013})},\ \Eprint {http://arxiv.org/abs/1311.2936}
  {arXiv:1311.2936 [cond-mat.str-el]} \BibitemShut {NoStop}%
\bibitem [{\citenamefont {Zaletel}\ and\ \citenamefont
  {Mong}(2012)}]{Zaletel12:MPS:Supplemental}%
  \BibitemOpen
  \bibfield  {author} {\bibinfo {author} {\bibfnamefont {M.~P.}\ \bibnamefont
  {Zaletel}}\ and\ \bibinfo {author} {\bibfnamefont {R.~S.~K.}\ \bibnamefont
  {Mong}},\ }\href {\doibase 10.1103/PhysRevB.86.245305} {\bibfield  {journal}
  {\bibinfo  {journal} {Physical Review B}\ }\textbf {\bibinfo {volume} {86}},\
  \bibinfo {pages} {245305} (\bibinfo {year} {2012})}\BibitemShut {NoStop}%
\bibitem [{\citenamefont {Wu}\ \emph {et~al.}(2014)\citenamefont {Wu},
  \citenamefont {Estienne}, \citenamefont {Regnault},\ and\ \citenamefont
  {Bernevig}}]{Wu14:MPS:Supplemental}%
  \BibitemOpen
  \bibfield  {author} {\bibinfo {author} {\bibfnamefont {Y.-L.}\ \bibnamefont
  {Wu}}, \bibinfo {author} {\bibfnamefont {B.}~\bibnamefont {Estienne}},
  \bibinfo {author} {\bibfnamefont {N.}~\bibnamefont {Regnault}}, \ and\
  \bibinfo {author} {\bibfnamefont {B.~A.}\ \bibnamefont {Bernevig}},\
  }\href@noop {} {} (\bibinfo {year} {2014}),\ \bibinfo {note} {in
  preparation}\BibitemShut {NoStop}%
\bibitem [{\citenamefont {Estienne}\ \emph {et~al.}(2014)\citenamefont
  {Estienne}, \citenamefont {Regnault},\ and\ \citenamefont
  {Bernevig}}]{Estienne14:Gaffnian:Supplemental}%
  \BibitemOpen
  \bibfield  {author} {\bibinfo {author} {\bibfnamefont {B.}~\bibnamefont
  {Estienne}}, \bibinfo {author} {\bibfnamefont {N.}~\bibnamefont {Regnault}},
  \ and\ \bibinfo {author} {\bibfnamefont {B.~A.}\ \bibnamefont {Bernevig}},\
  }\href@noop {} {} (\bibinfo {year} {2014}),\ \bibinfo {note}
  {unpublished}\BibitemShut {NoStop}%
\end{thebibliography}
\end{document}